\documentclass[aps,prl,twocolumn,longbibliography,superscriptaddress,notitlepage,floatfix]{revtex4-2}
\renewcommand{\figurename}{Fig.}
\usepackage{amsmath,amsfonts,amssymb,color,epsfig,graphics,graphicx,latexsym,theorem,url,multirow,diagbox}
\usepackage[colorlinks,colorlinks,citecolor=blue,linkcolor=blue,urlcolor=black]{hyperref}
\usepackage[utf8]{inputenc}
\usepackage{booktabs}
\usepackage[T1]{fontenc}
\usepackage{courier}
\usepackage{listings}
\usepackage{latexsym}
\usepackage{dcolumn}
\usepackage{comment}
\usepackage{times} 
\usepackage{algorithm}
\usepackage{algorithmic} 
\usepackage{braket}
\usepackage{bm}
\usepackage{lineno}
\usepackage{siunitx}
\renewcommand{\algorithmicrequire}{\textbf{Input}}  
\renewcommand{\algorithmicensure}{\textbf{Output}} 

\definecolor{blue}{rgb}{0,0,1}
\definecolor{red}{rgb}{1,0,0}
\definecolor{green}{rgb}{0,1,0}
\definecolor{sec_red}{RGB}{166, 40, 37}

\begin{document}
\title{Demonstration of low-overhead quantum error correction codes}
\affiliation{School of Physics, ZJU-Hangzhou Global Scientific and Technological Innovation Center, \\and Zhejiang Key Laboratory of Micro-nano Quantum Chips and Quantum Control, Zhejiang University, Hangzhou, China\\
$^{2}$Shanghai Qi Zhi Institute, Shanghai 200232, China\\
$^{3}$Center for Quantum Information, IIIS, Tsinghua University, Beijing 100084, China\\
$^{4}$School of Engineering and Applied Sciences, Harvard University, Cambridge, Massachusetts 02138, USA\\
$^{5}$School of Physics, Xi'an Jiaotong University, Xi'an 710049, China\\
$^{6}$International Research Centre MagTop, Institute of Physics, Polish Academy of Sciences, Aleja Lotnikow 32/46, PL-02668 Warsaw, Poland\\
$^{7}$Hefei National Laboratory, Hefei 230088, China}

\author{Ke Wang$^{1}$}\thanks{These authors contributed equally to this work.}
\author{Zhide Lu$^{2,3}$}\thanks{These authors contributed equally to this work.}
\author{Chuanyu Zhang$^{1}$}\thanks{These authors contributed equally to this work.}
\author{Gongyu Liu$^{1}$}
\author{Jiachen Chen$^{1}$}
\author{Yanzhe Wang$^{1}$}
\author{Yaozu Wu$^{1}$}
\author{Shibo Xu$^{1}$}
\author{Xuhao Zhu$^{1}$}
\author{Feitong Jin$^{1}$}
\author{Yu Gao$^{1}$}
\author{Ziqi Tan$^{1}$}
\author{Zhengyi Cui$^{1}$}	
\author{Ning Wang$^{1}$}
\author{Yiren Zou$^{1}$}
\author{Aosai Zhang$^{1}$}	
\author{Tingting Li$^{1}$}
\author{Fanhao Shen$^{1}$}
\author{Jiarun Zhong$^{1}$}
\author{Zehang Bao$^{1}$}
\author{Zitian Zhu$^{1}$}
\author{Yihang Han$^{1}$}
\author{Yiyang He$^{1}$}
\author{Jiayuan Shen$^{1}$}
\author{Han Wang$^{1}$}
\author{Jia-Nan Yang$^{1}$}
\author{Zixuan Song$^{1}$}
\author{Jinfeng Deng$^{1}$}
\author{Hang Dong$^{1}$}
\author{Zheng-Zhi Sun$^{3}$}
\author{Weikang Li$^{3}$}
\author{Qi Ye$^{3,4}$}
\author{Si Jiang$^{3}$}
\author{Yixuan Ma$^{3,5}$}
\author{Pei-Xin Shen$^{6}$}
\author{Pengfei Zhang$^{1}$}
\author{Hekang Li$^{1}$}
\author{Qiujiang Guo$^{1,7}$}
\author{Zhen Wang$^{1,7}$}
\email{2010wangzhen@zju.edu.cn}
\author{Chao Song$^{1,7}$}
\email{chaosong@zju.edu.cn}
\author{H. Wang$^{1,7}$}
\author{Dong-Ling Deng$^{3,2,7}$}\email{dldeng@mail.tsinghua.edu.cn}
\begin{abstract}{
\textbf{
Quantum computers hold the potential to surpass classical computers in solving complex computational problems. However, the fragility of quantum information and the error-prone nature of quantum operations make building large-scale, fault-tolerant quantum computers a prominent challenge. To combat errors, pioneering experiments have demonstrated a variety of quantum error correction codes. Yet, most of these codes suffer from low encoding efficiency, and their scalability is hindered by prohibitively high resource overheads. Here, we report the demonstration of two low-overhead quantum low-density parity-check (qLDPC) codes, a distance-4 bivariate bicycle code and a distance-3 qLDPC code, on our latest superconducting processor, Kunlun, featuring 32 long-range-coupled transmon qubits. Utilizing a two-dimensional architecture with overlapping long-range couplers, we demonstrate simultaneous measurements of all nonlocal weight-6 stabilizers via the periodic execution of an efficient syndrome extraction circuit. We achieve a logical error rate per logical qubit per cycle of (8.91 ± 0.17)\% for the distance-4 bivariate bicycle code with four logical qubits and (7.77 ± 0.12)\% for the distance-3 qLDPC code with six logical qubits. Our results establish the feasibility of implementing various qLDPC codes with long-range coupled superconducting processors, marking a crucial step towards large-scale low-overhead quantum error correction.}
}
\end{abstract}

\maketitle

\noindent
Quantum computing promises computational advantages in a wide range of applications \cite{nielsen2010quantum}, including factoring~\cite{Shor1995Scheme}, optimization~\cite{Farhi2001Quantum}, chemistry~\cite{Kandala2017Hardwareefficient}, finance~\cite{Herman2023Quantum}, and machine learning~\cite{Biamonte2017Quantum,DasSarma2019Machine}. Yet, for practical quantum computation to attain such advantages, errors on the physical level of the device need to be corrected so that large and deep circuits can be executed with desired accuracy. Quantum error correction provides a way to reach quantum fault tolerance through encoding logical qubits redundantly into many physical qubits, which enables one to diagnose and correct errors by repeatedly measuring error syndromes~\cite{Gottesman1997Stabilizer,Dennis2002Topological}.  Experimental demonstrations of different quantum error correction codes have been reported in diverse platforms, such as trapped ions~\cite{Fluhmann2019Encoding,Egan2021Faulttolerant,Ryan-Anderson2021Realization,DeNeeve2022Error,Paetznick2024Demonstration}, nitrogen-vacancy centers~\cite{Waldherr2014Quantum,Abobeih2022Faulttolerant,Chang2025Hybrid}, neutral atoms~\cite{Bluvstein2024Logical,Reichardt2024Logical}, and superconducting processors~\cite{Marques2022Logicalqubit,Zhao2022Realization,Krinner2022Realizing,GoogleQuantumAI2023Suppressing,Ni2023Beating,Sivak2023Realtime,Gupta2024Encoding,Lacroix2024Scaling,Caune2024Demonstrating,Eickbusch2024Demonstrating,GoogleQuantumAIandCollaborators2025Quantum}.

\begin{figure*}[t]
\centering
\includegraphics[width=1\textwidth]{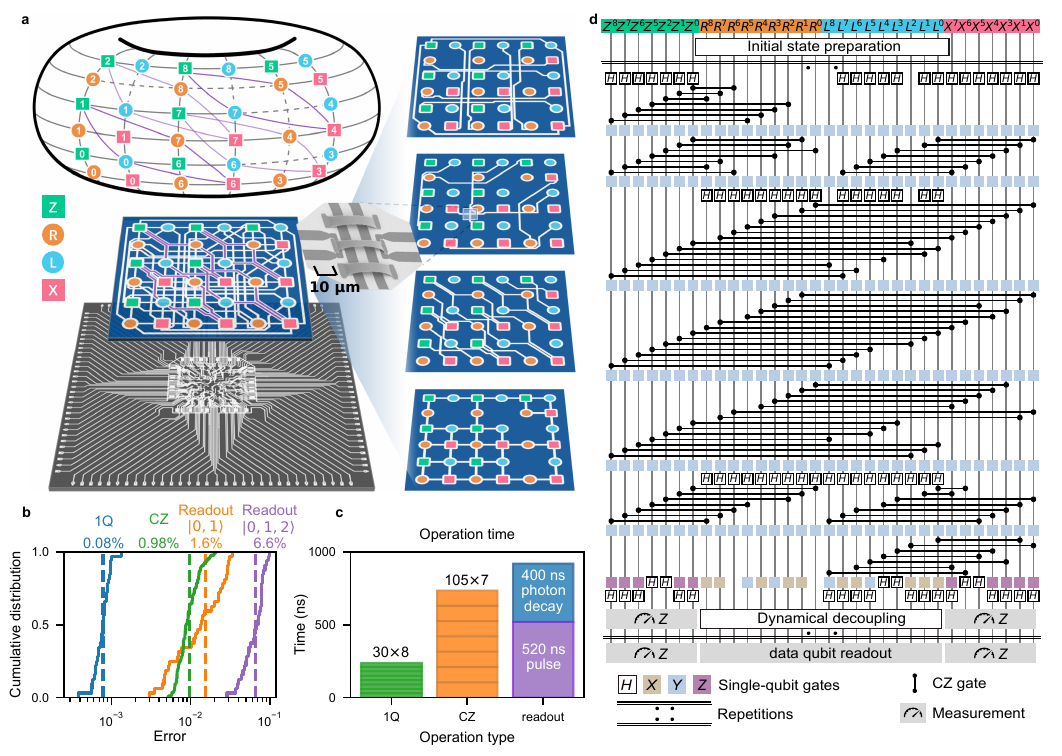}
\caption{
\textbf{Implementation of the bivariate bicycle code.}
\textbf{a}, Schematic of the Kunlun processor. 
(Upper left) Tanner graph of the bivariate bicycle (BB) code embedded into a torus, where relevant sites are indexed by color-coded shapes and numbers. 
Only part of long-range connections are shown.
(Lower left) Corresponding layout of the $32$-qubit superconducting quantum processor Kunlun that implements the $[[18,4,4]]$ BB code. 
The processor comprises $18$ data qubits (circles) categorized into $L$-type (blue) and $R$-type (orange), and $14$ check qubits (squares) for the $X$-type (red) and $Z$-type (green) stabilizer measurements. 
The purple lines highlight the couplers corresponding to the long-range connections in the Tanner graph.
(Right) The $84$ couplers are grouped into four sets according to their lengths, and a crossing region between two interleaved couplers is magnified to reveal the underlying microstructure.
\textbf{b}, Cumulative distributions of error probabilities measured on Kunlun. Blue: Pauli errors for parallel single-qubit gates; green: Pauli errors for parallel CZ gates; orange (purple): assignment errors for simultaneous two (three) -state measurements. Dashed vertical lines denote the mean values.
\textbf{c}, Operation times of quantum operations in a single error correction cycle. The total cycle duration is \SI{1895}{\nano\second}, including eight \SI{30}{}-\SI{}{\nano\second} single-qubit gate layers, seven \SI{105}{}-\SI{}{\nano\second} CZ gate layers, a \SI{520}{}-\SI{}{\nano\second} readout pulse, and a \SI{400}{}-\SI{}{\nano\second} photon ring-down time. The measurement of the data qubit in the last cycle is performed with an \SI{890}{}-\SI{}{\nano\second} readout pulse.
\textbf{d}, Syndrome measurement circuit of the $[[18,4,4]]$ BB code. The data qubits (labeled $L^i$ and $R^i$) are initially prepared in product states.
The first syndrome cycle of stabilizer measurements projects the data qubits into a logical state, and the subsequent cycles extract stabilizers for error correction. Each syndrome cycle contains seven layers of CZ gates interleaved with single-qubit gates to simultaneously extract $X$-type and $Z$-type stabilizers, and ends with the measurement of the $X$-type and $Z$-type check qubits (labeled $X^i$ and $Z^i$).
In the final cycle, the data qubits are measured to obtain both stabilizer values and the information of the logical state. We insert additional Pauli $X$ and $Y$ gates between CZ layers and implement dynamical decoupling during the measurement of the check qubits to protect the data qubits from dephasing.
}
\label{fig:Fig1}
\end{figure*}

A crucial aspect of quantum error correction codes concerns their encoding efficiency. Many existing codes fall short in this aspect, including the surface code~\cite{Kitaev2003Faulttolerant,Bravyi1998Quantum,Dennis2002Topological} that for $20$ years was the leading approach towards fault-tolerant quantum computing. Indeed, in the surface code a single logical qubit is encoded into a $d\times d$ square of physical qubits, leading to an encoding rate scaling as $1/d^2$ and vanishing in the large $d$ limit. This compromises its scalability to hundreds or more logical qubits, despite the fact that small-scale surface codes with a single logical qubit have been demonstrated by several groups~\cite{Marques2022Logicalqubit,Zhao2022Realization,Krinner2022Realizing,GoogleQuantumAI2023Suppressing,GoogleQuantumAIandCollaborators2025Quantum,Eickbusch2024Demonstrating,Caune2024Demonstrating}. An intriguing way out of this dilemma is to consider more general quantum low-density parity-check (qLDPC) codes~\cite{Breuckmann2021Quantum}, which would achieve quantum fault tolerance with a much higher encoding efficiency. In particular, a special family of qLDPC codes, known as bivariate bicycle (BB) codes, has recently been proposed and shown to exhibit high encoding efficiency  \cite{Bravyi2024Highthreshold}. Under the circuit-based noise model, these codes merit an error threshold close to $0.7\%$, which is comparable to that of the surface code. In addition, they support a low-depth syndrome measurement circuit, an efficient decoding algorithm, and a fault-tolerant protocol for implementing logical gates. Despite these merits and the fact that hardware requirements for realizing BB codes are relatively mild compared to that for several other qLDPC codes such as hypergraph product~\cite{Tillich2014Quantum} or quantum Tanner codes~\cite{Panteleev2022Asymptotically,Leverrier2022Quantum}, the implementation of BB codes is exceedingly more challenging than that of the surface code and thus has evaded experiment so far. In fact, BB codes pose a more stringent demand on designing a new category of superconducting processors with higher connectivity and more freedom in control greatly surpassing the conventional two-dimensional architecture~\cite{GoogleQuantumAIandCollaborators2025Quantum,Kim2023Evidence,Jin2025Observation}, in which remotely separated physical qubits must possess precisely programmable couplings to allow for direct gate operations.

Here, we overcome these difficulties and report the demonstration of a distance-4 BB code and a distance-3 qLDPC code on our latest superconducting processor, Kunlun, with $32$ long-range-coupled transmon qubits (Fig.~\ref{fig:Fig1}). We design and fabricate our Kunlun processor to bear a torus connectivity topology, which is adapted to planar architectures by integrating multi-length tunable couplers for distant qubits separated by length scales up to $6.5$~mm (Fig.~\ref{fig:Fig1}\textbf{a}).
Our device supports programmable interconnects among qubits with a maximum degree of six, meeting the connectivity requirement for implementing these codes. The average fidelities for the parallel single- and two-qubit gates are $99.95\%$ and $99.22\%$, respectively. For the distance-4 BB code, we encode four logical qubits into 18 physical qubits and use the remaining $14$ qubits to serve as check qubits. We repeatedly execute an efficient syndrome measurement circuit over multiple cycles to extract error syndromes, which are then decoded using the belief propagation with an ordered statistics decoder (BP-OSD)~\cite{Roffe2020Decoding,Panteleev2021Degenerate} extended to the circuit-based noise model. We obtain a  logical error rate per logical qubit per cycle of $(8.91\pm 0.17)\%$. 
The encoding rate for our implemented BB code is $1/8$, which offers approximately a fourfold reduction of the encoding overhead compared with the distance-$4$ surface code. 
To implement the distance-$3$ qLDPC code, we encode six logical qubits on $18$ data qubits and use $12$ check qubits for stabilizer measurements. By decoding the error syndromes, we obtain a logical error rate per logical qubit per cycle of $(7.77\pm0.12)\%$. 
Our work paves the way for implementing diverse low-overhead qLDPC codes with superconducting processors.

\begin{figure*}[t]
\centering
\includegraphics[width=1\textwidth]{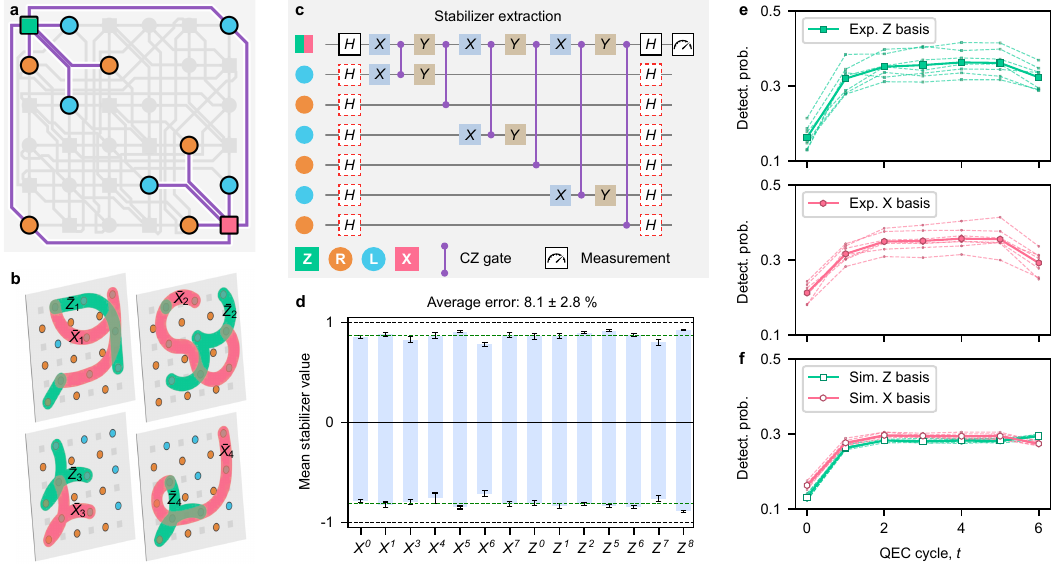}
\caption{
\textbf{Stabilizer extraction and error detection in the bivariate bicycle code.}
\textbf{a}, A $Z$-type (green squares) or $X$-type (red squares) check qubit is coupled to six data qubits (circles), including three $L$-type (blue) and three $R$-type (orange), through both local and long-range couplers.
\textbf{b},
Logical Pauli operators $\bar{Z}$ and $\bar{X}$ for the four logical qubits encoded in the distance-$4$ BB code.
Each logical $\bar{Z}$ (green shaded) or $\bar{X}$ (red shaded) is the tensor product of the Pauli $Z$ or $X$ operators acting on specific sets of the data qubits.
\textbf{c},
Quantum circuit for extracting a weight-six $X$-type or $Z$-type stabilizer.
The circuit applies sequential CZ gates between the check qubit and each of the six data qubits. The Hadamard gates marked by red dashed squares are applied on the data qubits only when measuring the $X$-type stabilizer.
Additional Pauli $X$ and $Y$ gates are inserted to mitigate qubit dephasing errors.
\textbf{d}, Measured mean stabilizer values for the $14$ check qubits. For each stabilizer, the positive and negative bar heights correspond to averages over the $32$ basis states of even and odd parities, respectively.
The overall averages across all $14$ stabilizers are indicated by green dashed lines, from which the average error can be calculated.
\textbf{e}, Error detection probability over seven syndrome cycles for logical $Z$-basis and $X$-basis preservation experiments (more than $40,000$ experimental instances after leakage rejection).
Dashed lines represent individual stabilizers and the solid line represents the average over all stabilizers.
\textbf{f}, Similar to \textbf{e}, using a simulation based on the circuit-based depolarizing noise model.
}
\label{fig:Fig2}
\end{figure*}

\vspace{.4cm}
\noindent\textbf{\large{}The bivariate bicycle code} \\
\noindent Bivariate bicycle codes are a family of Calderbank–Shor–Steane quantum stabilizer codes constructed from bivariate polynomials~\cite{Bravyi2024Highthreshold}. We use the notation $[[n,k,d]]$ to describe an error correction code that encodes $k$ distance-$d$ logical qubits into the joint entangled state of $n$ data qubits. 
For our experimental demonstration, we have designed a BB code with parameters $[[18,4,4]]$. The four logical qubits are defined by a set of logical Pauli operators $\left\{\bar{X}_1, \bar{Z}_1, \ldots, \bar{X}_4, \bar{Z}_4\right\}$. This distance-$4$ BB code can be described by a Tanner graph embedded on a torus with periodic boundary conditions (Fig.~\ref{fig:Fig1}\textbf{a}), where vertices correspond either to data (for encoding logical qubits) or check qubits (for measuring error syndromes).
An edge connecting a check qubit and a data qubit indicates that the corresponding check operator acts non-trivially on the data qubit.
Each check operator is geometrically non-local and acts non-trivially on six data qubits\textemdash four local and two long-range. All $X$-type (or $Z$-type) check operators can be obtained through horizontal and vertical translations of a single one across the grid.

By partitioning the check matrices into the left and right parts (Methods), the $18$ data qubits can be categorized into $L$-type and $R$-type, accordingly. Each data qubit is labeled either as $L^{i}$ or $R^{i}$, where $i$ ranges from $1$ to $9$.
The check qubits are labeled as $X^{j}$ (measuring $X$-type stabilizers) or $Z^{j}$ (measuring $Z$-type stabilizers) with $j$ ranging from $1$ to $9$.
We note that the direct construction of BB codes includes a total of $18$ check operators, but only $14$ of them are independent. We thus remove four redundant check operators and retain only the independent ones to reduce experimental complexities.

\vspace{.4cm}
\noindent\textbf{\large{}Experimental setup and framework}

\noindent Our experiments are performed on a flip-chip superconducting quantum processor with $32$ frequency-tunable transmon qubits arranged in a square lattice (Fig.~\ref{fig:Fig1}\textbf{a}).
The average lifetime of all physical qubits is measured to be $T_1=$ \SI{41.8}{\micro\second}.
To implement the distance-$4$ BB code, we use $18$ qubits as data qubits and the remaining $14$ as check qubits.
The distance-$3$ qLDPC code with parameters $[[18,6,3]]$ is implemented using $18$ data qubits and $12$ check qubits. Compared to the BB code, the check qubits $X^5$ and $Z^5$ and the couplers connecting them to the data qubits are not used (Methods).
The processor possesses $84$ tunable couplers with lengths ranging from $1$ mm to $6.5$ mm, offering flexible interconnects between check and data qubits with lattice distances up to five.
In particular, each check qubit is coupled to six data qubits via both local and long-range couplers.
To realize the torus connectivity topology of the degree-$6$ Tanner graph on a planar architecture, we introduce air-bridges with numbers up to $15$ into the couplers to allow for multi-coupler crossovers. See Supplementary Information for more device information.

In our experiment, the stabilizer measurement circuits are compiled using Hadamard and CZ gates, with additional Pauli $X$ and $Y$ gates inserted to reduce the dephasing errors.
Single-qubit gates are realized by applying resonant microwave pulses with a duration of \SI{30}{\nano\second}, and have an average fidelity of $99.95\%$ as determined by simultaneous cross-entropy benchmarking among all qubits.
The $84$ pairs of CZ gates have a duration of \SI{105}{\nano\second} each, including two {\SI{5}{}-\SI{}{\nano\second}-long} idling times at the beginning and the end, and exhibit an average fidelity of $99.22\%$ (Fig.~\ref{fig:Fig1}\textbf{b}). 
We characterize CZ gate fidelity using simultaneous cross-entropy benchmarking with sets of $7$ or $14$ gates executed in parallel, as employed in our realization of the distance-4 BB code.
We observe similar performances of CZ gates among qubit pairs interconnected by either local or long-range couplers.
To execute the syndrome measurement circuit for multiple cycles, it is crucial to reduce the readout times of the check qubits, as the data qubits can suffer from decoherence when measuring the check qubits.
We realize a readout duration of \SI{920}{\nano\second} for the check qubits, including a \SI{520}{}-\SI{}{\nano\second}-long readout pulse and another \SI{400}{}-\SI{}{\nano\second}-long idling time for photon decaying (Fig.~\ref{fig:Fig1}\textbf{c}).
As a result, we obtain an average readout error of $1.6\%$ for simultaneous two-computational-state readout and of $6.6\%$ for simultaneous three-state readout, which discriminates between the two computational states and a leakage state.

With high-precision, long-range two-qubit gates, we characterize the measurements of individual stabilizers (Fig.~\ref{fig:Fig2}).
For each stabilizer, we sequentially prepare its six coupled data qubits in all $2^6=64$ basis states, and then execute the gate sequence for extracting a weight-$6$ stabilizer (Fig.~\ref{fig:Fig2}\textbf{a, c}) for $3,000$ repetitions per basis.
We separately compute the mean stabilizer measurement outcomes for the $32$ basis states of odd parity and the other $32$ states of even parity.
As shown in Fig.~\ref{fig:Fig2}\textbf{d}, the average of measured $14$ stabilizer values are $0.868$ for even parity and $-0.808$ for odd parity. From these values, we estimate the average experimental error in measuring weight-$6$ stabilizers to be $(8.1\pm2.8)\%$.
We integrate all individual stabilizer measurements into a syndrome measurement circuit, enabling simultaneous extraction of all stabilizer values.
As illustrated in Fig.~\ref{fig:Fig1}\textbf{d}, the full cycle of the syndrome measurement circuit includes qubit state initialization, seven layers of non-overlapping CZ gates, eight layers of single-qubit gates, and qubit readout. 
To protect qubits from dephasing, we insert single-qubit gates between CZ gates and apply dynamical decoupling gates during qubit idles.

With all elements in place, we implement the distance-$4$ BB code for logical state preservation. The logical Pauli operators corresponding to the four encoded logical qubits are illustrated in Fig.~\ref{fig:Fig2}\textbf{b}.
In a single experimental run, we begin with encoding a four-logical-qubit state. To encode a logical $Z$ basis state, we first initialize each data qubit randomly into either $|0\rangle$ or $|1\rangle$. Subsequently, we execute one cycle of stabilizer measurements, setting the four-logical-qubit state into $|\phi_1\phi_2\phi_3\phi_4\rangle_{\text{L}}$. Each $\phi_i$ is determined by the parity of the corresponding logical operator $\bar{Z}_i$: $\phi_i=0$ for even parity and $\phi_i=1$ for odd parity.
Preparing logical $X$ basis states follows a similar procedure.
After encoding, we execute multiple consecutive syndrome cycles, recording the stabilizer outcomes to construct the error syndromes (see Methods).
In the final cycle, we directly measure all data qubits to extract both stabilizer parity information and logical measurement outcomes.
We decode the error syndromes to infer the most likely error configuration and then apply corrections in post-processing.
An experimental instance is considered successful if the corrected logical state agrees with the initial encoded logical state; otherwise, a logical error is recorded.
To evaluate the logical performance, we perform up to six consecutive syndrome cycles. For each cycle $t=1,\ldots,6$, we collect data from over $40,000$ experimental instances (excluding instances in which leakage is detected) and decode the error syndromes to obtain the logical error probability.
We fit the logical error probability as a function of the number of executed syndrome cycles to extract the logical error per cycle.

\begin{figure*}[t]
\centering
\includegraphics[width=0.9\textwidth]{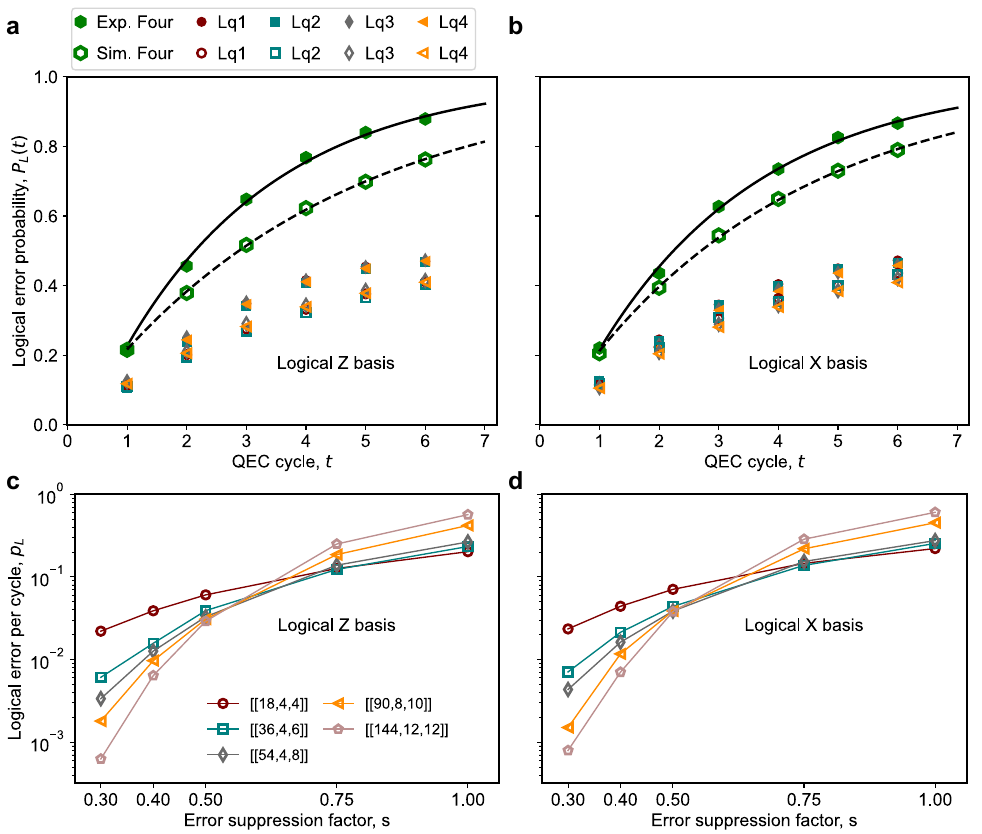}
\caption{
\textbf{Logical error rate per cycle and performance prediction.} 
\textbf{a} and \textbf{b},
Logical error probability $P_L(t)$ (filled symbols: experiment; empty symbols: simulation) versus the number of executed cycles, for logical $Z$ basis state preservation and logical $X$ basis state preservation, respectively.
Each individual data point represents more than $40,000$ experimental or simulated instances.
Logical error probabilities for each of the four logical qubits are also shown.
Solid line: fit to the experimental data from $t = 1$ to $6$; Dashed line: fit to the simulation.
The fit is performed using the least-squares method on $\log(1- P_L(t))$ versus $t$.
Lq, logical qubit; QEC, quantum error correction.
\textbf{c} and \textbf{d},
Simulated logical error rate per cycle $p_L$ of bivariate bicycle codes, as functions of code distance and an error suppression factor $s$ on the current physical error rates exhibited in Fig.~\ref{fig:Fig1}\textbf{c}, semilog plot. $s=1$ corresponds to the physical error rates in the current device.
}
\label{fig:Fig3}
\end{figure*}

\vspace{.4cm}
\noindent\textbf{\large{}Error detection} \\
\noindent Physical errors in the syndrome measurement circuit can be detected by comparing stabilizer values across consecutive syndrome cycles, based on which the error syndromes are constructed. Any difference between a stabilizer's value and its corresponding value in the preceding cycle indicates an error.
To characterize the distribution of errors while implementing the distance-$4$ BB code, we present the detection probabilities across seven syndrome cycles, including the initial cycle for logical state preparation.
The results shown in Fig.~\ref{fig:Fig2}\textbf{e} are based on data aggregated from more than $40,000$ experimental instances, excluding any experimental runs in which leakage is detected.
The average detection probability is $0.319$ for $Z$-type stabilizers and $0.318$ for $X$-type stabilizers.
The lower detection probabilities observed in the initial and final cycles arise from boundary effects~\cite{GoogleQuantumAI2021Exponential}.
In the initial cycle, stabilizer values are compared to the expected values determined by the initial state preparation of data qubits. In the final cycle, stabilizer values are directly obtained by data qubit readout. Both processes introduce fewer errors compared to a full syndrome cycle.
The experimental results for the distance-$3$ qLDPC code are presented in Extended Data Fig.~\ref{fig:Detection_prob_18_6_3}.
The average detection probability is $0.312$ for both $X$-type and $Z$-type stabilizers, closely matching the values observed in the distance-$4$ BB code. This is anticipated as both codes measure weight-$6$ stabilizers and are implemented on the same Kunlun processor.

To help understand the experimental results, we simulate the BB code experiment using the circuit-based noise model, based on the physical errors exhibited in Fig.~\ref{fig:Fig1}\textbf{b}.
The simulation results are shown in Fig.~\ref{fig:Fig2}\textbf{f}. We find that the average detection probability is $0.259$ for $Z$-type stabilizers and $0.270$ for $X$-type stabilizers, both being slightly smaller than the corresponding experimental values. We attribute this discrepancy to the fact that these numerical simulations have not accounted for other experimental imperfections, such as crosstalk effects.
The simulation for the distance-$3$ qLDPC code yields an average detection probability that closely matches the BB code, as shown in Extended Data Fig.~\ref{fig:Detection_prob_18_6_3}.

In comparison to surface codes~\cite{Krinner2022Realizing,GoogleQuantumAI2023Suppressing}, the error detection probabilities in the BB code and distance-$3$ qLDPC code are higher.
In addition to factors such as the fidelity of simultaneous CZ gates and readout, we attribute the higher detection probabilities to weight-$6$ stabilizers in the BB code.
Errors detected by a weight-$6$ stabilizer may originate from any of the six coupled data qubits, thus resulting in a higher detection probability than that of weight-$4$ or weight-$2$ stabilizers typically used in surface codes.

\vspace{.4cm}
\noindent\textbf{\large{}Decoding and logical error rates}\\
To examine the logical performance of the distance-$4$ BB code and the distance-$3$ qLDPC code, we decode the error syndromes and correct the inferred errors in post-processing.
For decoding, we use the BP-OSD decoder extended to the circuit-based depolarizing noise model.
After decoding and applying correction, we compute the logical error probabilities $P_L(t)$ as a function of the number of executed cycles $t$, based on more than $40,000$ experimental instances (after leakage rejection).

The results are shown in Fig.~\ref{fig:Fig3}\textbf{a},\textbf{b}.
To extract the logical error rate per cycle $p_{L}$, we fit the data using the relation $1-P_L\propto(1-p_{L})^{t}$.
We then estimate the logical error per qubit per cycle as $\varepsilon_k = 1-(1-p_{L})^{1/k}$, where $k$ is the number of logical qubits encoded in the implemented code.
For the distance-$4$ BB code, we obtain a logical error rate per logical qubit per cycle of $(9.15 \pm 0.27)\%$ for logical $Z$-basis state preservation and $(8.67 \pm 0.21)\%$ for logical $X$-basis state preservation, with an average value of $(8.91\pm 0.17)\%$. The error bar indicates one standard deviation for statistical and fit uncertainty.
In terms of encoding efficiency, implementing four distance-$4$ surface codes would require $124$ physical qubits in total\textemdash nearly four times the number of qubits in the distance-$4$ BB code implementation.
For the distance-$3$ qLDPC code (Extended data Fig.~\ref{fig:Logical_error_18_6_3}), similar fitting yields a logical error per logical qubit per cycle of $(7.79 \pm 0.12)\%$ for $Z$ bases and $(7.75 \pm 0.21)\%$ for $X$ bases, with an average of $(7.77 \pm 0.12)\%$.

To estimate the logical performance of BB codes in future devices with improved performance, we simulate the experiment of BB codes with distances ranging from $4$ to $12$ under varying physical error rates. The specific code parameters of BB codes used in the simulations are provided in Extended Data Tab.~\ref{tab:BB_parameter}.
We introduce a suppression factor $s$ to represent that each physical error rate used in the simulation is scaled to $s$ times the corresponding value in the current device (Fig.~\ref{fig:Fig1}\textbf{b}).
The simulation results (Fig.~\ref{fig:Fig3}\textbf{c},\textbf{d}) indicate that at the current noise level ($s=1$), the logical error rate increases with code distance, suggesting that codes with larger distance do not yet benefit from logical error suppression. 
In contrast, when physical error rates are sufficiently reduced, increasing the code distance would lead to a suppression of logical error rates.
The transition between these two regimes occurs at approximately $s=0.5$.
It is worth noting that this analysis relies on a simple circuit-based noise model. A more accurate evaluation should account for additional experimental imperfections, such as crosstalk and leakage, to better characterize real devices.

\vspace{.4cm}
\noindent\textbf{\large{}Discussion}\\
\noindent In summary, we have experimentally demonstrated two low-overhead qLDPC codes with our new  Kunlun superconducting processor. These codes offer a three to four-times reduction of the encoding overhead compared to the surface codes that encode the same number of logical qubits with the same code distance. The success of our experiments relies crucially on high-fidelity CZ gates between distant transmon qubits achieved on the Kunlun processor by using overlapping long-range couplers. The controllability and long-range connectivity demonstrated in our experiment clearly manifest the potential of realizing other qLDPC codes with the superconducting platform, which we leave for future investigations. 

Several challenges remain. First, the suppression of logical error requires that the physical error rate be below a critical threshold. In our current experiments, this has not been achieved yet, and the logical error rate remains larger than that of the physical qubits for both the distance-$4$ BB and distance-$3$ qLDPC codes. 
To attain and go beyond the break-even point, further technical improvements are demanded, including the realization of two-qubit gates with even higher fidelities and on qubit pairs separated by even larger distances, faster and more accurate readout of check qubits~\cite{Jeffrey2014Fast,Heinsoo2018Rapid,Sunada2022Fast,Spring2024Fast,Sank2025System}, and a more efficient leakage removal method~\cite{McEwen2021Removing,Miao2023Overcoming}.
Second, practical fault-tolerant quantum computing demands that a universal logical gate set can be implemented in a fault-tolerant fashion, which is exceedingly challenging and has so far evaded experimental demonstration for both the surface code and qLDPC codes considered in the current work. Third, it is also of crucial importance to scale up our experiments to larger low-overhead qLDPC codes and demonstrate exponential suppression of logical errors with increasing code distance. This would mark another key step towards large-scale, fault-tolerant quantum computing. 

\vspace{.1cm}
\makeatother
\bibliographystyle{naturemag}
\bibliography{qldpc}

\clearpage
\noindent\textbf{\large{}Methods}

\noindent\textbf{Code construction} \\
We construct the $[[18,4,4]]$ BB code following ref.~\cite{Bravyi2024Highthreshold}.
We define two binary matrices: $x = S_{l} \otimes I_m $ and $y = I_{l} \otimes S_m$, where $I_{m}$ is an identity matrix of size $m \times m$ and $S_{l}$ is a cyclic shift matrix of size $l \times l$, with the $i$-th row having a single non-zero entry in the column ($i+1$ mod $l$).
Using these matrices, we define two polynomials $A(x,y)= x^{a_1} + y^{a_2} + y^{a_3}$ and $B(x,y)= y^{b_1} + x^{b_2} + x^{b_3}$.
The $X$-type and $Z$-type check matrices are then given by $H_X  =[A | B]$ and $H_Z =\left[B^\top | A^\top \right]$, where the bar $|$ denotes horizontal matrix concatenation, and $\top$ denotes matrix transposition.
The matrix $H_X$ is composed of the left part $A$ and the right part $B$, which determine the connections between $X$-type check qubits and the $L$-type and $R$-type data qubits, respectively.
Similarly, the connections between $Z$-type check qubits and the $L$-type and $R$-type data qubits are determined by the left part $B^\top$ and the right part $A^\top$.
To construct the BB code with parameters $[[18,4,4]]$ in our experiment, we select $l=3$ and $m=3$, with $[a_1,a_2,a_3]=[1,0,2]$ and $[b_1,b_2,b_3]=[1,0,2]$. Based on these parameters, we derive the corresponding check matrices and compute the code parameters $k=4$ and $d=4$.
Among the $18$ check operators, only $14$ are independent. to reduce experimental complexity, we remove four redundant check operators measured by check qubits $X^2$, $X^8$, $Z^3$, and $Z^4$, and retain only the independent ones.
To construct the qLDPC code with parameters $[[18,6,3]]$, we further remove check operators measured by check qubits $X^5$ and $Z^5$. The logical Pauli operators for both the BB code $[[18,4,4]]$ and the qLDPC code $[[18,6,3]]$ are provided in Supplementary Information Sec. IA.

\vspace{2mm}
\noindent\textbf{Data analysis} \\
Similar to ref.~\cite{Krinner2022Realizing}, we convert the measurement outcomes of check qubits into stabilizer values (Supplementary Information Sec. IIIA), since the check qubits are not reset between syndrome cycles.
For the first cycle, the stabilizer values are directly given by the measurement outcomes of check qubits.
For the final cycle, the stabilizer values are extracted from the outcomes of the data qubit readout.
For the middle cycles, the measurement outcomes of check qubits do not directly represent the stabilizer values since the check qubit states are not reset to $|0\rangle$ after executing the preceding cycle. 
Instead, the stabilizer values in cycle $t$ are determined by the parity difference between the measurement outcomes of check qubits in cycles $t$ and $t-1$.
Thus, if consecutive cycles yield the same outcomes, the stabilizer value is $1$; otherwise, the stabilizer value is $0$.

\vspace{2mm}
\noindent\textbf{Extracting logical error per cycle} \\
With the obtained logical error probabilities $P_L(t)$ over varying cycles $t$, we fit data points $\left\{\left(t, P_L\left(t\right)\right)\right\}_{t=1}^{6}$ using the functional form 
\begin{equation}
P_L(t)=1-A(1-p_L)^{t}
\end{equation}
to extract the logical error $p_L$ per cycle. To linearize this model, we define $a = \log A$ and $b = \log(1-p_L)$, transforming the model into 
\begin{equation}
y(t) = \log(1-P_L(t)) = a + b \cdot t .
\end{equation}
We perform a linear fitting using the least-squares method to determine the best-fit parameters $a^*$ and $b^*$, as well as the standard error $\sigma(b)$.
Finally, we exponentiate the fitted parameters to recover the original model parameters $A = e^{a^*}$ and $p_{L}=1-e^{b^*}$.
To estimate the error bar for $p_{L}$, we apply error propagation to $b^*$, yielding
\begin{equation}
\sigma(p_L) = \left|\frac{d p_L}{d b}\right| \sigma(b) = e^b \sigma(b) = (1-p_L)\sigma(b) .
\end{equation}

\vspace{2mm}
\noindent\textbf{Numerical simulation} \\
We numerically simulate the experiment using a circuit-based depolarizing noise model, where each operation is followed by a randomly injected Pauli error with a specific probability (Supplementary Information Tab.~S1). 
The operations include single-qubit gates (Hadamard gates), CZ gates, idle operations, dynamical decoupling, check-qubit measurements, and final-cycle readout of data qubits. 
Each operation in the syndrome measurement circuit is assumed to experience independent Pauli noise, with error probabilities set to the mean physical error rates shown in Fig.~\ref{fig:Fig1}\textbf{b}.
During dynamical decoupling, the error probability for each data qubit is taken to be $[1- \exp(-\tau/T)]/2$, where $\tau$ is the duration of the check qubit readout and $T$ is the coherence time. 
Note that we do not explicitly consider errors from qubit state initialization, since qubits are initialized only at the beginning of each experimental run and check qubits are not periodically reset. Actually, the effects of faulty initialization can be absorbed into the subsequent Hadamard and CZ gates.

With all elements in place, we simulate the logical error performance of the distance-$4$ BB code and distance-$3$ qLDPC code as functions of the number of cycles $t=1,\ldots,6$.
For each $t$, we perform $40,000$ simulation repetitions.
After decoding the error syndromes, we compute the logical error probabilities $P_L(t)$ versus $t$, and then fit the data to extract the simulated logical error per cycle.
The fitting for the distance-$4$ BB code (Fig.~\ref{fig:Fig3}\textbf{a},\textbf{b}) yields a logical error per qubit per cycle of $(5.82\pm 0.04)\%$ and $(6.45\pm 0.06)\%$ for logical $Z$ bases and $X$ bases, respectively.
For the distance-$3$ qLDPC code, we obtain a logical error per qubit per cycle of $(4.88\pm 0.01)\%$ (logical $Z$ bases) and $(5.39\pm 0.08)\%$ (logical $X$ bases), as shown in Extended Data Fig.~\ref{fig:Logical_error_18_6_3}.

\vspace{2mm}
\noindent\textbf{Logical performance projection} \\
To estimate the logical performance of BB codes with distances ranging from $4$ to $12$ (Extended data Tab.~\ref{tab:BB_parameter}), we simulate each code under varying error suppression factors $s$. For each $s$ and $d$, we simulate $t=6$ cycles of syndrome measurements, decode the error syndromes, and compute the logical error probability $P_L(6)$. We then estimate the logical error rate per cycle as $p_L=1-(1-P_L(6))^{1/6}$~\cite{Bravyi2024Highthreshold}.

\vspace{2mm}
\noindent\textbf{Decoding} \\
We decode the error syndromes using the BP-OSD decoder~\cite{Roffe2020Decoding,Panteleev2021Degenerate} extended to the circuit-based depolarizing noise model (Supplementary Information Sec. IC).
We first simulate each scenario involving exactly one faulty operation in the syndrome measurement circuit and record the resulting error syndromes.
Next, we group together faulty operations that yield the same error syndrome and compute the sum of their associated error probabilities (Supplementary Information, Tab.~S1). 
After this preparation stage, the BP-OSD decoder is applied to the error syndromes to infer the most likely error configuration by seeking a solution with the highest reliability in the formulated optimization problem.
For the decoding, we configure the BP-OSD decoder with a maximum of $10,000$ iterations, a variable scaling factor, and the ``OSD-CS'' method with a search depth of $7$.

\vspace{.6cm}
\noindent\textbf{\large{}Data availability}
The data presented in the figures and that support the other findings of this study will be made publicly available for download on Zenodo upon publication.

\vspace{.6cm}
\noindent\textbf{\large{}Code availability}
The data analysis and numerical simulation codes will be made publicly available for download on Zenodo upon publication.

\vspace{.5cm}
\noindent\textbf{Acknowledgement} 
We thank Dong Yuan and Wenjie Jiang for helpful discussions. 
{The device was fabricated at the Micro-Nano Fabrication Center of Zhejiang University.  We acknowledge support from the Innovation Program for Quantum Science and Technology (grant nos.~2021ZD0300200 and 2021ZD0302203), the National Natural Science Foundation of China (grant nos.~12174342, 92365301, 12274367, 12322414, 12274368, 12075128, 12404570, 12404574, T2225008, and T24B2002), the Shanghai Qi Zhi Institute Innovation Program SQZ202318, the National Key R\&D Program of China (grant no.~2023YFB4502600), and the Zhejiang Provincial Natural Science Foundation of China (grant nos.~LDQ23A040001, LR24A040002). 
Z.-Z.S., W.L., and D.-L.D. are supported in addition by Tsinghua University Dushi Program.
P.-X.S. acknowledges support from the European Union's Horizon Europe research and innovation programme under the Marie Skłodowska-Curie Grant Agreement No.~101180589 (SymPhysAI), the National Science Centre (Poland) OPUS Grant No.~2021/41/B/ST3/04475, and the ``MagTop'' project (FENG.02.01-IP.05-0028/23) carried out within the ``International Research Agendas'' programme of the Foundation for Polish Science co-financed by the European Union under the European Funds for Smart Economy 2021-2027 (FENG).
Views and opinions expressed are however those of the author(s) only and do not necessarily reflect those of the European Union or the European Research Executive Agency. Neither the European Union nor the granting authority can be held responsible for them.
}

\vspace{.3cm}
\noindent\textbf{Author contributions}  K.W. and C.Z.~carried out the experiments under the supervision of C.S.~and H.W.. J.C. and Y.Z.W. designed the device, and J.C. and H.L. fabricated the device supervised by H.W.. Z.W. designed the control and measurement electronics. Z.L. designed the error correction codes and performed the numerical simulations under the supervision of D.-L.D.. Z.L., Z.-Z.S., W.L., Q.Y., S.J., Y.M., P.S., and D.-L.D.~conducted the theoretical analysis. All authors contributed to the experimental set-up, the discussions of the results, and the writing of the manuscript.

\vspace{.3cm}
\noindent\textbf{Competing interests}  All authors declare no competing interests.

\vspace{.5cm}
\renewcommand{\figurename}{Extended Data Fig.}
\renewcommand{\tablename}{Extended Data Tab.}
\setcounter{figure}{0}
\setcounter{table}{0}

\begingroup
\setlength{\tabcolsep}{12.5pt} 
\renewcommand{\arraystretch}{2.3} 
\begin{center}
\begin{table*}
\begin{tabular}{c|c|c|c|c}
\hline \hline
$[[n,k,d]]$ & Encoding rate & $\ell, m$ & $A$ & $B$  \\
\hline
$[[18,4,4]]$ & $1/9$ & $3,3$ & $x+y^0+y^2$  & $y+x^0+x^2$   \\
\hline
$[[36,4,6]]$ & $1/18$ & $6,3$ & $x^1+y^0+y^1 $ & $x^2+y^2+y^3$  \\
\hline
$[[54,4,8]]$ & $$1/27$$ & $9,3$ & $x^1+y^0+y^2$ & $y^1+x^6+x^5$  \\
\hline
$[[90,8,10]]$ & $1/22.5$ & $3,15$ & $x^0+y^1+y^5$ & $y^3+x^1+x^2$   \\
\hline
$[[144,12,12]]$ & $1/24$ & $12,6$ & $x^3+y^1+y^2$ & $y^3+x+x^2$  \\
\hline \hline
\end{tabular}
\caption{\textbf{Parameters of bivariate bicycle codes used for logical performance projection.} 
A code with parameters $[[n,k,d]]$ encodes $k$ distance-$d$ logical qubits into $n$ data qubits. 
In this construction, a total of $2n$ check qubits are used, including $n$ data qubits and $n$ check qubits. The encoding rate is evaluated as $k/2n$.
}
\label{tab:BB_parameter}   
\end{table*}
\end{center}
\endgroup

\begin{figure*}[htb]
\includegraphics[width=0.9\textwidth]{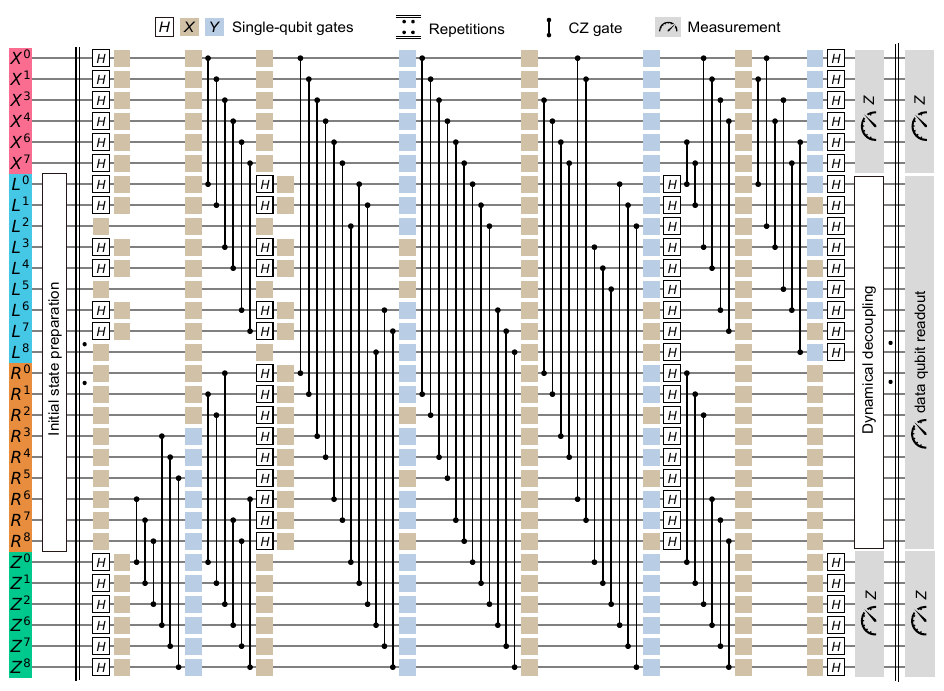}
\caption{ 
\textbf{Quantum circuit used to initialize and repeatedly error correct the six logical qubits of qLDPC code $\boldsymbol{[[18,6,3]]}$.}
The $X$-type checks and $Z$-type checks are marked with red and green colors, respectively. The data qubits of $L$ and $R$ type are marked with blue and orange colors, respectively.
}
\label{fig:circuit_18_6_3}
\end{figure*}

\begin{figure*}[htb]
\includegraphics[width=0.95\textwidth]{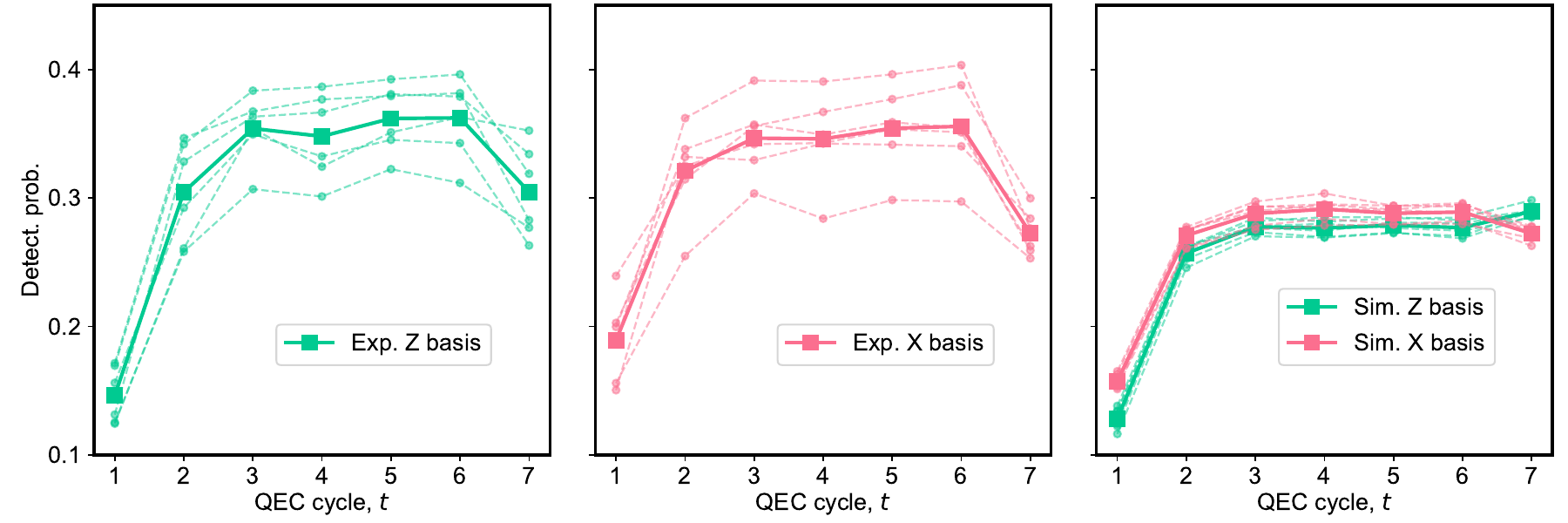}
\caption{ 
\textbf{Detection probability for stabilizers over seven cycles for the qLDPC code with parameters $\boldsymbol{[[18,6,3]]}$.}
Each data point is obtained from over $40,000$ experimental instances.
The dotted lines indicate the detection probability for each individual stabilizer, and the solid line shows the average detection probability across all stabilizers of the $Z$-type or $X$-type.
}
\label{fig:Detection_prob_18_6_3}
\end{figure*}

\begin{figure*}[htb]
\includegraphics[width=0.95\textwidth]{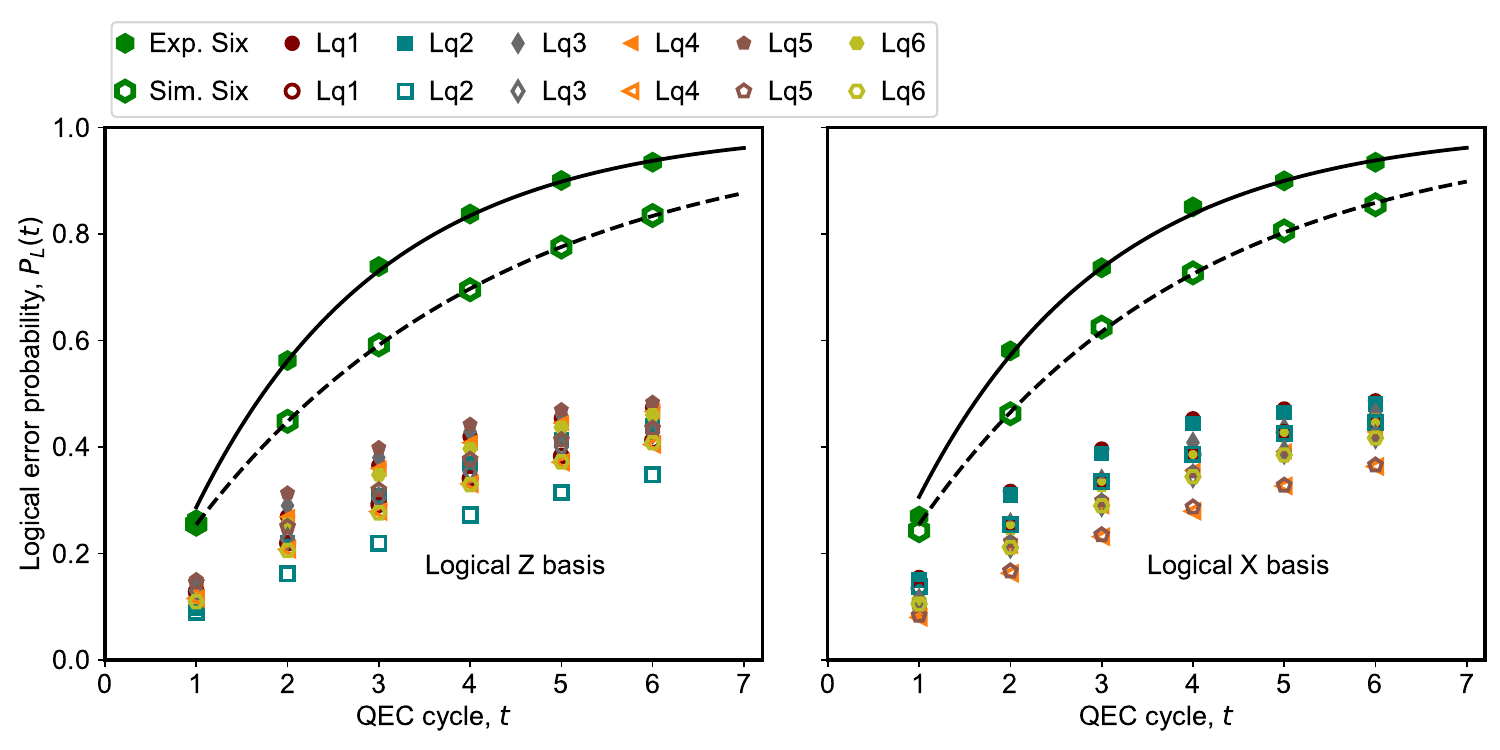}
\caption{ 
\textbf{Accumulated logical error probabilities as a function of the number of cycles for the qLDPC code with parameters $\boldsymbol{[[18,6,3]]}$.}
The logical error probability for the six logical qubits and for each individual logical qubit is shown. Each data point represents over $40,000$ experimental instances after leakage rejection.
The solid line corresponds to an exponential fit to the data.
}
\label{fig:Logical_error_18_6_3}
\end{figure*}

\clearpage
\newpage 
\onecolumngrid

\newcommand{\mcol}{\multicolumn{2}{c}}
\newcommand{\cfill}{\cellcolor[HTML]{44cef6}}
\newcommand{\cfillo}{\cellcolor[HTML]{f05654}}
\setcounter{MaxMatrixCols}{10}
\hypersetup{urlcolor=blue}

\xdef\presupfigures{\arabic{figure}}
\newcommand{\figpath}{./figures}
\renewcommand{\algorithmicrequire}{\textbf{Input}}  
\renewcommand{\algorithmicensure}{\textbf{Output}} 

\setcounter{figure}{0}
\setcounter{table}{0}
\renewcommand{\theequation}{S\arabic{equation}}
\renewcommand{\thefigure}{S\arabic{figure}}
\renewcommand{\thetable}{S\arabic{table}}
\renewcommand{\figurename}{Fig.}
\renewcommand{\tablename}{Tab.}

\begin{center} 
{\large \bf Supplementary Information for \\ ``Demonstration of low-overhead quantum error correction codes''}
\end{center}

\maketitle

\section{Theoretical details}

Quantum error correction (QEC) is the cornerstone of fault-tolerant quantum computing~\cite{Preskill1998Reliable,Terhal2015Quantum,Campbell2017Roads}. Traditional QEC codes, such as the surface code, face scalability challenges due to their high resource overhead~\cite{Fowler2012Surface}.
Recently, the focus has shifted toward more general quantum low-density parity-check (qLDPC) codes~\cite{Breuckmann2021Quantum}. These codes may offer high encoding efficiency and efficient resource scaling, making them compelling candidates for enabling practical, large-scale quantum computing.
Our experiment focuses on a particular family of qLDPC codes known as bivariate bicycle (BB) codes~\cite{Bravyi2024Highthreshold}. 
In this section, we will discuss the construction of BB codes, the syndrome measurement circuit for stabilizer extraction, and the decoder.

\subsection{Bivariate bicycle code construction}
We describe a QEC code with parameters $[[n, k, d]]$, where $k$ logical qubits are encoded into $n$ data qubits, and the code distance $d$ indicates that any logical error spans at least $d$ data qubits.

Let $\mathcal{P}_n = \left\{ i^{\alpha}\left\{ I, X, Y, Z \right\}^{\otimes n}  | \alpha \in\{0,1,2,3\}  \right\}$ be the $n$-qubit Pauli group, where $I$ is the identity operator and $X, Y, Z$ are Pauli matrices.
Quantum stabilizer codes are defined by an Abelian subgroup  $\mathcal{S}$ of the Pauli group $\mathcal{P}_n$; each element of $\mathcal{S}$ is called a stabilizer. 
The code space is the common $+1$-eigenspace of all stabilizers in $\mathcal{S}$.
To construct an $[[n, k, d]]$ stabilizer code, one typically chooses a stabilizer group $\mathcal{S}$ generated by $n-k$ independent stabilizers. Consequently, the dimension of the code space is $2^k$, thereby encoding $k$ logical qubits.

We denote the parity-check matrix of an $[[n, k, d]]$ stabilizer code by $H\in \mathbb{F}_{2}^{(n-k)\times 2n}$. Each row of $H$ is a binary string $(x_1, x_2,\ldots,x_n | z_1, z_2,\ldots,z_n)$, 
which corresponds to a stabilizer Pauli operator given by:
\begin{equation}
\bigotimes_{i=1}^n X^{x_i} \bigotimes_{i=1}^n Z^{z_i} .
\end{equation}
Calderbank-Shor-Steane (CSS) codes form a specific subclass of stabilizer codes characterized by stabilizer generators that consist entirely of either $X$-type or $Z$-type Pauli operators.
The parity-check matrix for a CSS code can be expressed in the block-diagonal form as:
\begin{equation}
H_{\text{CSS}} = \left(\begin{array}{cc}
H_X & \mathbf{0} \\
\mathbf{0} & H_Z
\end{array}\right)  ,
\end{equation}
where $H_X$ and $H_Z$ are the $X$-type and $Z$-type check matrices, respectively.
The requirement that the stabilizer group is Abelian translates into the condition: $H_X H^{\top}_Z = 0$.

Before formally defining BB codes, we introduce some matrices.
Let $I_{m}$ be the $m \times m$ identity matrix, and let $S_{l}$ be the $l \times l$ cyclic shift matrix, defined as:
\begin{equation}
S_l=\sum_{i=0}^{l-1}|i\rangle\langle i+1 \ \text{mod} \ l| .
\end{equation}
For example,  
\begin{equation*}
S_4=\left[\begin{array}{llll}
0 & 1 & 0 & 0 \\
0 & 0 & 1 & 0 \\
0 & 0 & 0 & 1 \\
1 & 0 & 0 & 0 \\     
\end{array} \right]  .
\end{equation*}
Using $I_{m}$ and $S_{l}$, we define two binary matrices $x = S_{l} \otimes I_m $ and $y = I_{l} \otimes S_m$. 
Following Ref.~\cite{Bravyi2024Highthreshold}, a BB code is a CSS-type code with parity-check matrices determined by two polynomials $A(x,y)$ and $B(x,y)$ of the following form:
\begin{equation}
\begin{aligned}
    A(x, y) &= x^{a_1} + y^{a_2} + y^{a_3} \\
    B(x, y) &= y^{b_1} + x^{b_2} + x^{b_3}  ,
\end{aligned}
\end{equation}
where the addition and multiplication between matrices are performed in module $2$. Since $x^l = y^m = I_{lm}$, the integers $a_1, b_2, b_3$ are constrained to the range $[0, l-1]$, while $b_1, a_2, a_3$ are selected from $[0, m-1]$.
The $X$-type and $Z$-type check matrices of the BB code are then defined as:
\begin{equation}
\begin{aligned}
H_X & =[A | B] \\
H_Z & =\left[B^\top | A^\top \right] ,
\label{eq:check_matrices}
\end{aligned}
\end{equation}
where the notation ``$|$'' indicates horizontal stacking of matrices, and ``$\top$'' denotes matrix transposition.
The notation ``$|$'' partitions the data qubits into the $L$-type and $R$-type (see the main text).
It can be verified that $H_X H_Z^{\top}=AB+BA=0 \ (\operatorname{mod} 2)$, satisfying the CSS condition.
Note that each row of $H_X$ or $H_Z$ contains exactly six non-zero entries, determined by the three terms in $A(x,y)$ and the three terms in $B(x,y)$. This implies that each check operator acts nontrivially on six data qubits.
Since $H_X$ and $H_Z$ have $2lm$ columns, the BB code contains $2n=4lm$ physical qubits, including $n$ data qubits and $n$ check qubits.

According to Lemma $1$ in ref.~\cite{Bravyi2024Highthreshold}, the BB code with parameters $[[n, k, d]]$ defined by $A$ and $B$ satisfy the following relations:
\begin{equation}
\begin{aligned}
    k &= n - 2\times \operatorname{rank}(H_X) = 2\times \operatorname{dim}(\operatorname{ker}(A) \cap \operatorname{ker}(B) )  \\
    d &= \min \left\{|\mathbf{v}|: \mathbf{v} \in \operatorname{ker}\left(H_X\right) \backslash \mathrm{rs}\left(H_Z\right)\right\} .
\label{eq:lemma 1}
\end{aligned}   
\end{equation}
Here, $|\mathbf{v}|$ denotes the number of non-zero entries in $\mathbf{v}$; $\operatorname{ker}(A)$ denotes the nullspace of $A$, consisting of all vectors $v$ satisfying $Av = 0 \ (\operatorname{mod} 2)$; $\operatorname{rs}(H_Z)$ is the row space of $H_Z$, spanned by its rows.

For convenience in the following discussion, we adopt the same notation as in the main text: $X^i$, $Z^i$, $L^i$ and $R^i$ represent the $i$-th $X$-type check, $Z$-type check, $L$-type data, $R$-type data, respectively, where $i=1,2, \ldots, n/2$.
The connectivity between $L^i$ and checks is determined by the $i$-th column of $H_X$ and $H_Z$, while the connectivity of $R^i$ is determined by the $(i+n/2)$-th column of $H_X$ and $H_Z$.

\begin{figure}[t]
\includegraphics[width=0.8\textwidth]{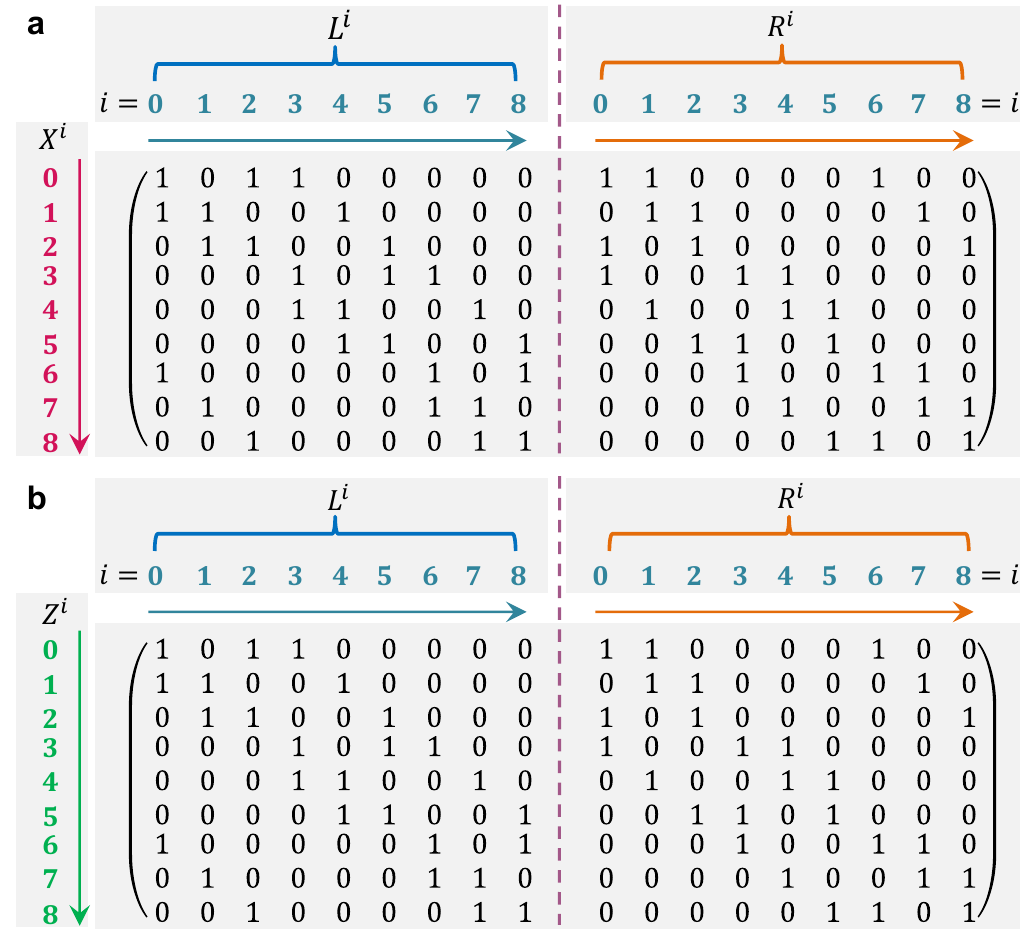}
\caption{\textbf{Check matrices of the BB code with parameters $\boldsymbol{[[18,4,4]]}$.} 
\textbf{a,} The $X$-check matrix $H_X$. Each row represents an $X$-type check operator, acting non-trivially on the data qubits at the position indicated by non-zero entries. The column indices correspond to the data qubit labels, ranging from $L^0$ to $L^8$ in the left block and from $R^0$ to $R^8$ in the right block.
Each row contains six non-zero entries, indicating that each $X$-check operator acts non-trivially on six data qubits.
\textbf{b,} The $Z$-check matrix $H_Z$. Similarly, each row has six non-zero entries, with three in the left block and three in the right block.
}
\label{fig:check_matrices}
\end{figure}

For our experimental demonstration, we design a BB code with parameters $[[n=18, k=4, d=4]]$. 
First, we select $l=3$ and $m=3$, which gives $n = 2lm = 18$. By definition, we set $x = S_{3} \otimes I_{3} $ and $y = I_{3} \otimes S_{3}$.
Next, we choose $a_1=1, a_2=0, a_3=2$ and $b_1=1, b_2=0, b_3=2$, leading to $A = x^{1} + y^{0} + y^{2}$ and $B = y^{1} + x^{0} + x^{2}$. Based on these definitions, we derive the corresponding check matrices $H_X$ and $H_Z$.
We then compute the code parameters using Equation~\ref{eq:lemma 1}, obtaining $k=4$ and $d=4$. Thus, the encoding rate is $k/(2n)=1/9$.

\begin{figure}[t]
\includegraphics[width=0.75\textwidth]{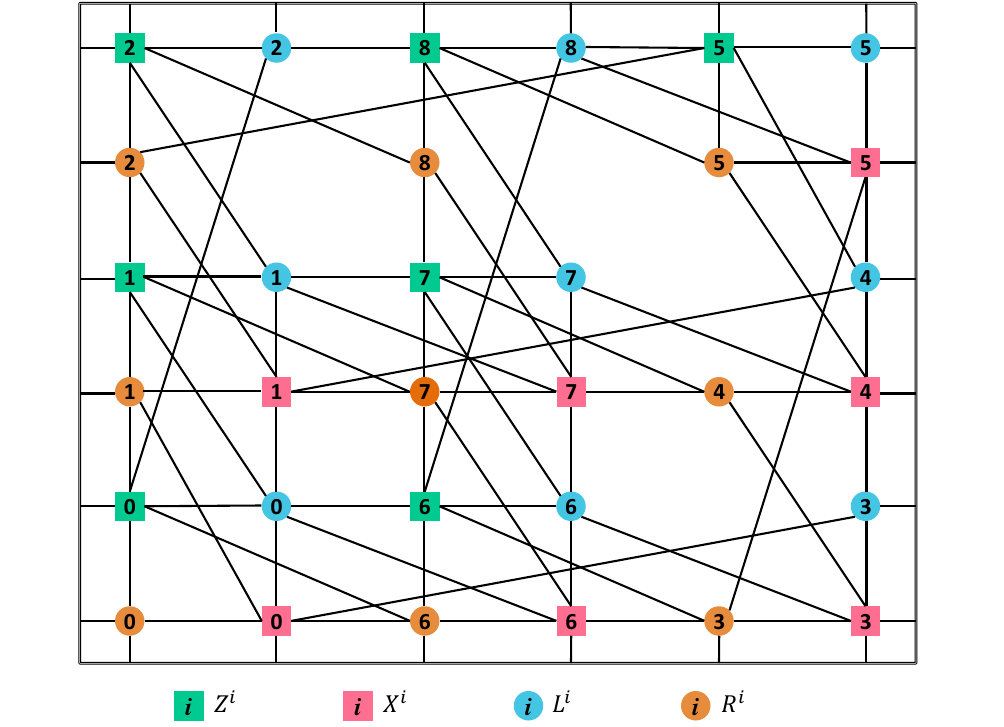}
\caption{\textbf{Tanner graph of the BB code with parameters $\boldsymbol{[[18, 4, 4]]}$ embedded into a toric layout.}
In the Tanner graph, the leftmost qubit is connected to the rightmost qubit, and the top qubit is connected to the bottom qubit, reflecting the toric structure. 
The redundant check operators $X^2$, $X^8$, $Z^3$, and $Z^4$ are removed.
Each edge connects a data qubit and a check qubit. Data qubits of $L$-type and $R$-type are depicted by blue and orange circles, respectively. Check qubits of $X$-type and $Z$-type are shown as red and green squares.}
\label{fig:lay_out}
\end{figure}

In Fig.~\ref{fig:check_matrices}, we present the complete entries of the check matrices $H_X$ and $H_Z$.
We observe that in our case, $H_X = H_Z$, which results from our choice of parameters. Specifically, the following holds:
\begin{equation}
\begin{aligned}
    A &= x^{1} + y^{0} + y^{2}  \\
      &= x^{-2} + x^{0} + y^{-1}  \\
      &= (x^{2})^\top + x^{0} + (y^{1})^\top  \\
      &= ( (y^{1}) + x^{0} + (x^{2}) )^\top    \\
      &= B^\top   .
\end{aligned}
\end{equation}
Similarly, $B = A^\top$, and according to Equation~\ref{eq:check_matrices}, it follows that $H_X = H_Z$. 
In addition, we find that $\operatorname{rank}(H_X)=\operatorname{rank}(H_Z)=7$, indicating that among the nine $X$-type stabilizers, only seven are independent. Thus, we remove two redundant $X$-type and two redundant $Z$-type check operators.
The permitted selections of $X$-type check operators whose removal do not change the code parameters are listed below: 
\begin{equation*}
\begin{array}{llllll}
\big\{X^0, X^1\big\}, &  \big\{X^0, X^2\big\},  & \big\{X^0, X^3\big\}, & \big\{X^0, X^4\big\}, & \big\{X^0, X^6\big\}, & \big\{X^0, X^8\big\},  \\
\big\{X^1, X^2\big\}, & \big\{X^1, X^4\big\}, & \big\{X^1, X^5\big\}, & \big\{X^1, X^6\big\}, & \big\{X^1, X^7\big\}, & \big\{X^2, X^3\big\}, \\
\big\{X^2, X^5\big\}, & \big\{X^2, X^7\big\}, & \big\{X^2, X^8\big\}, & \big\{X^3, X^4\big\}, & \big\{X^3, X^5\big\}, & \big\{X^3, X^6\big\}, \\
\big\{X^3, X^7\big\}, & \big\{X^4, X^5\big\}, & \big\{X^4, X^7\big\}, & \big\{X^4, X^8\big\}, & \big\{X^5, X^6\big\}, & \big\{X^5, X^8\big\}, \\
\big\{X^6, X^7\big\}, & \big\{X^6, X^8\big\}, & \big\{X^7, X^8\big\}. & & &
\end{array}
\end{equation*}
Similar results hold for $Z$-type check operators.
In our implementation, we remove $X$-type check operators $X^2$ and $X^8$, and $Z$-type check operators $Z^3$ and $Z^4$.
Based on $H_X$ and $H_Z$, we present the Tanner graph of the BB code $[[18,4,4]]$ in Fig.~\ref{fig:lay_out}.
In the Tanner graph, each vertex represents either a data qubit or a check qubit. An edge connects the $i$-th check vertex to the $j$-th data vertex if the $i$-th check operator acts non-trivially on the $j$-th data qubit.
We observe that the BB code $[[18,4,4]]$ exhibits a two-dimensional toric layout. However, unlike the toric code with geometrically local check operators, the check operators in the BB code are not geometrically local: each check qubit acts non-trivially on six data qubits, with at least two long-range couplings.

We select a set of logical Pauli-$X$ and Pauli-$Z$ operators for the four encoded logical qubits:
\begin{itemize}
\item $\bar{X}_1 = X\left(  L^0,L^2,L^4, L^5,L^6,L^7 \right)  $ \qquad $\bar{Z}_1 = Z\left(  L^0,L^2,L^3, L^4,L^8,R^0 \right)  $
\item $\bar{X}_2 = X\left(  L^1,L^2,L^3, L^4,L^6,L^8 \right)   $ \qquad $\bar{Z}_2 = Z\left(  L^0,L^2,L^4, L^5,L^6,L^7 \right)  $
\item $\bar{X}_3 = X\left(  L^0,L^1,L^6, R^0 \right)   $ \qquad \qquad \quad $\bar{Z}_3 = Z\left(  L^1,L^2,L^7, R^1 \right)  $
\item $\bar{X}_4 = X\left(  L^0,L^1,L^4,L^5,L^6, R^1 \right)  $ \qquad $\bar{Z}_4 = Z\left( L^0,L^1,L^6, R^0 \right)  $
\end{itemize}
Here, the notation 
\begin{equation}   
 X\left( L^{l_{1}},L^{l_{2}},\ldots, R^{r_{1}},R^{r_{2}}, \ldots \right) 
\end{equation}
denotes the tensor product of Pauli-$X$ operators acting on the data qubits $L^{l_{1}}, L^{l_{2}},\ldots$ and $R^{r_{1}}, R^{r_{2}},\ldots$. A similar notation is used for the logical Pauli-$Z$ operators.

Based on the BB code $[[18,4,4]]$, we construct a qLDPC code with parameters $[[18,6,3]]$ by removing two check operators $X^5$ and $Z^5$. Removing these two check operators increases the number of encoded logical qubits from four to six while reducing the code distance from four to three.
For the encoded six logical qubits, we choose a set of logical Pauli-$X$ and Pauli-$Z$ operators as follows:

\begin{itemize}
\item $\bar{X}_1 = X\left(L^0, L^1, L^3, L^5, L^6\right)$ \qquad 
$\bar{Z}_1 = Z\left(L^0, L^3, L^4, L^5, L^8\right)$
\item $\bar{X}_2 = X\left(L^1, L^2, L^3, L^4, L^7\right)$ \qquad 
$\bar{Z}_2 = Z\left(L^0, L^3, L^8, R^0, R^1\right)$
\item $\bar{X}_3 = X\left(L^0, L^2, L^4, L^5, L^8\right)$ \qquad 
$\bar{Z}_3 = Z\left(L^0, L^5, L^8, R^0, R^2\right)$
\item $\bar{X}_4 = X\left(L^3, L^5, R^0\right)$ \qquad\qquad\quad 
$\bar{Z}_4 = Z\left(L^1, L^4, L^6, L^8, R^0, R^1, R^2\right)$
\item $\bar{X}_5 = X\left(L^3, L^4, R^1\right)$ \qquad\qquad\quad 
$\bar{Z}_5 = Z\left(L^0, L^1, L^2, L^4, L^7, L^8, R^2\right)$
\item $\bar{X}_6 = X\left(L^4, L^5, R^2\right)$ \qquad\qquad\quad 
$\bar{Z}_6 = Z\left(L^2, L^4, R^1\right)$
\end{itemize}

\begin{table}
\centering
\renewcommand{\arraystretch}{1.8}  
\begin{tabular}{c|c}
\hline \hline
\textbf{Component operation} & \textbf{Error probability} \\
\hline
H   & $8.0 \times 10^{-4}$   \\ \hline
I   & $3.5 \times 10^{-3}$   \\ \hline
DD ($p_{\text{D}}^{(X)}$)  &  $1.09 \times 10^{-2}$ \\ \hline
DD ($p_{\text{D}}^{(Z)}$) & $1.59 \times 10^{-2}$   \\ \hline  
CZ  & $9.8 \times 10^{-3}$   \\ \hline
Meas. (check)   & $4.03 \times 10^{-2}$   \\ \hline
Readout. (data)   & $3.29 \times 10^{-2}$   \\
\hline \hline
\end{tabular}
\caption{\textbf{Component error probabilities used in the decoder and classical simulation.}}
\label{tab:error_rates}
\end{table}

\subsection{Syndrome measurement circuit}
\begin{figure}[t]
\includegraphics[width=1\textwidth]{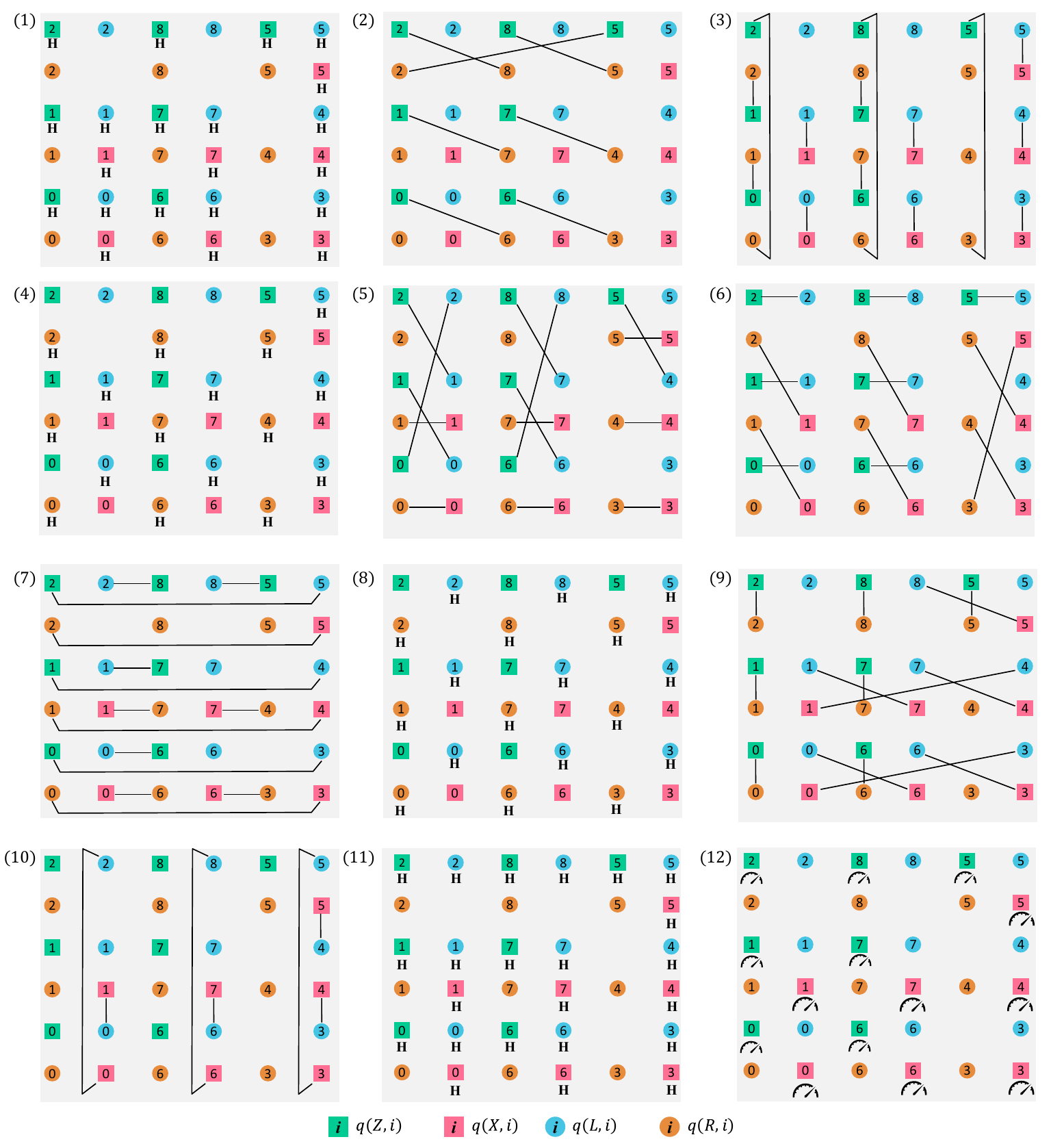}
\caption{\textbf{The full cycle of the syndrome measurement circuit for the BB code with parameters $\boldsymbol{[[18,4,4]]}$.}
Initially, each check qubit is prepared in the $|0\rangle$ state.
The circuit includes seven layers of CZ gates, as illustrated in ref.~\cite{Bravyi2024Highthreshold}.
In each round, an edge connecting a data qubit and a check qubit represents a CZ gate applied to them.
In the final round, the check qubits are measured on the Pauli-$Z$ basis.
}
\label{fig:SM_circuit}
\end{figure}

To detect and correct physical errors, we periodically measure all stabilizers and record the corresponding error syndromes.
The circuit used to extract the value of a single stabilizer to the corresponding $X$-type or $Z$-type check qubit is exhibited in Fig.~2\textbf{c} of the main text.
In our experiment, we implement an efficient syndrome measurement circuit (proposed in ref.~\cite{Bravyi2024Highthreshold}) that simultaneously extracts all stabilizer values.
For the BB code $[[18,4,4]]$, the full syndrome cycle is presented in Fig.~\ref{fig:SM_circuit}.
A full syndrome cycle contains seven layers of non-overlapping CZ gates, and notably, the number of layers of CZ gates remains constant as BB codes scale up.
It has been demonstrated that seven layers of CZ gates represent the minimal configuration required across all alternative syndrome measurement circuit designs~\cite{Bravyi2024Highthreshold}.
Note that the CNOT gates in the syndrome measurement circuit proposed in ref.~\cite{Bravyi2024Highthreshold} are replaced by a combination of CZ and Hadamard gates, as the CZ gate is directly implementable in the experiment.

\subsection{Decoding}
To decode the error syndromes obtained in the experiment, we use a classical belief propagation with ordered statistics decoding (BP-OSD) decoder~\cite{Roffe2020Decoding,Bravyi2024Highthreshold} to infer the most likely physical errors on data qubits given the observed error detections.
Decoding succeeds if the guessed error is equivalent to the actual error modulo a product of check operators; in this case, applying the inverse of the inferred error to the data qubits restores the original logical state.
Conversely, decoding fails if the inferred error differs from the actual error by a non-trivial logical operator.
In this subsection, we first introduce a circuit-based depolarizing noise model that simulates our BB code experiment and then apply the BP-OSD decoder to this noise model.

The circuit-based noise model assumes that Pauli noise is applied independently to each component operation in the syndrome measurement circuit with a certain prior probability. 
These component operations include single-qubit gates (Hadamard gates), CZ gates, idle gates, dynamical decoupling (DD), measurements of check qubits, and the final cycle readout of data qubits.
For each operation, the noise effect is characterized as follows:
\begin{itemize}
    \item \textbf{Hadamard gates:} Each Hadamard gate has an error probability $p_{\text{H}}$. A faulty Hadamard gate is modeled as an ideal Hadamard gate followed by a single-qubit Pauli error. This Pauli operator is randomly sampled from $\{X, Y, Z\}$ with equal probability $p_{\text{H}}/3$.
    \item \textbf{Idle gates:} Each idle gate has an error probability $p_{\text{I}}$. The error channel for an idle gate involves randomly applying one of the Pauli errors $\{X, Y, Z\}$ with probability $p_{\text{I}}/3$.
    \item \textbf{CZ gates:} The error probability for each CZ gate is $p_{\text{cz}}$. A faulty CZ gate is modeled by inserting a Pauli operator after an ideal CZ gate. The Pauli operator is randomly chosen from the
    $15$ possible non-identity two-qubit Pauli errors, each occurring with identical probability $p_{\text{cz}}/15$.
    \item \textbf{DD:} Each DD operation has an error probability $p_{\text{D}}$ and is modeled in a manner similar to an idle gate.
    \item \textbf{Measurement and data qubit readout:} The error probability for each check qubit measurement is $p_{\text{M}}$, and for each data qubit readout in the final cycle is $p_{\text{F}}$. A faulty measurement (or readout) produces the outcome opposite to that of an ideal measurement. As the readout error does not alter the check qubit state itself, it is modeled by inserting a Pauli $X$ flip both before and after the readout operation.
\end{itemize}
Note that the noise model does not explicitly consider qubit state initialization since our experiment initializes qubits only at the beginning and does not periodically reset check qubits. Therefore, the effects of faulty initialization are absorbed into the error probabilities for Hadamard or CZ gates rather than treated as a separate error source.

The probabilities $\{p_{\text{H}},p_{\text{I}},p_{\text{cz}},p_{\text{D}},p_{\text{M}},p_{\text{F}}\}$ listed in Tab.~\ref{tab:error_rates} are taken from the experimentally characterized error rates in Fig.~1C of the main text.
For simplicity, we assume that all operations of the same type share the same error probability, even though the actual values may vary at different positions in the syndrome measurement circuit. For instance, while $p_{\text{cz}}$ may differ among individual CZ gates and $p_{\text{M}}$ may vary across different check qubits, we use a constant $p_{\text{cz}}$ for all CZ gates and a constant $p_{\text{M}}$ for all check qubits.
The probability $p_{\text{D}}$ is estimated using the experimental coherence time $T_{1}$ and $T_{2}$. Specifically, the Pauli $X$ and Pauli $Z$ error rates are given by
\begin{equation}
\begin{aligned}
&p_{\text{D}}^{(X)}=\frac{1}{2}-\frac{1}{2} \exp \left(-\frac{\tau}{T_1}\right), \\
&p_{\text{D}}^{(Z)}=\frac{1}{2}-\frac{1}{2} \exp \left(-\frac{\tau}{T_2}\right),
\end{aligned}
\end{equation}
where $\tau$ denotes the DD time, i.e., the duration for measuring the check qubits.
Accounting for leakage, we estimate $p_{\text{M}}$ and $p_{\text{F}}$ as follows.
We begin with an experimentally determined three-state confusion matrix,
\begin{equation}
\begin{bmatrix}
q_{00} & q_{01} & q_{02} \\
q_{10} & q_{11} & q_{12} \\
q_{20} & q_{21} & q_{22} ,
\end{bmatrix}
\end{equation}
where $q_{ij}$ is the probability of the qubit being in $|i\rangle$ but read out as $|j\rangle$. Here, $i, j \in\{0,1,2\}$, with $|2\rangle$ denoting a leakage state.
We assume a leakage probability $\eta$, and let the probability of the qubit being in the computational basis ($|0\rangle$ or $|1\rangle$) be $2\beta$, so that $2\beta+\eta=1$.
After rejecting any detected leakage, we normalize the remaining outcomes to obtain a two-state confusion matrix:
\begin{equation}
\left[\begin{array}{cc}
q^\prime_{00} & q^\prime_{01} \\
q^\prime_{10} & q^\prime_{11}
\end{array}\right]
=
\left[\begin{array}{cc}
\frac{\beta q_{00}+\eta q_{20} / 2}{\beta q_{00}+\eta q_{20} / 2+\beta q_{01}+\eta q_{21} / 2} & \frac{\beta q_{01}+\eta q_{21} / 2}{\beta q_{00}+\eta q_{20} / 2+\beta q_{01}+\eta q_{21} / 2} \\
\frac{\beta q_{10}+\eta q_{20} / 2}{\beta q_{10}+\eta q_{20} / 2+\beta q_{11}+\eta q_{21} / 2} & \frac{\beta q_{11}+\eta q_{21} / 2}{\beta q_{10}+\eta q_{20} / 2+\beta q_{11}+\eta q_{21} / 2}
\end{array}\right] .
\end{equation}
The readout error rate is estimated as $(q^\prime_{10}+q^\prime_{01})/2$. Under the chosen parameters ($\eta=0.05$ and $\beta=0.475$), we obtain $p_{\text{M}}=4.03\times10^{-2}$ and $p_{\text{F}}=3.29\times10^{-2}$ (see Tab.~\ref{tab:error_rates}).

The BP-OSD decoder begins with a preparation step.
In this stage, we simulate all scenarios in which a single faulty operation occurs within the syndrome measurement circuit and track the resulting error syndromes. Note that the simulation is efficiently carried out using the stabilizer formalism, as the circuit involves solely Clifford operations.
Once the preparation is completed, the BP-OSD decoder is then applied to the measured error syndromes from the experiment to infer the most likely error configuration.

We consider running the syndrome measurement circuit for BB code $[[18,4,4]]$ for $t$ QEC cycles and denote the overall circuit by $\mathcal{U}$.
The circuit $\mathcal{U}$ contains the following components:
\begin{itemize}
    \item $6t(n-4)=84t$ CZ gates (accounting for the removal of four check qubits),
    \item $14t$ check qubit measurement operations,
    \item $n=18$ data qubit readout in the final cycle,
    \item $78t$ Hadamard gates,
    \item $18(t-1)$ DD operations, and additional idle gates.
\end{itemize}
Each CZ gate has $15$ possible faulty realizations; each Hadamard, DD, or idle gate has three faulty realizations; and each check qubit measurement or data qubit readout has a single faulty realization.
In total, there are $M$ possible faulty realizations of $\mathcal{U}$, which we denote as $\{ \mathcal{U}_1, \mathcal{U}_2, \ldots, \mathcal{U}_M \}$. Each faulty realization occurs with a probability determined by the error rates $\{p_{\text{H}}/3,p_{\text{I}}/3,p_{\text{cz}}/15,p_{\text{D}}/3,p_{\text{M}},p_{\text{F}}\}$ for the respective operations.
We simulate each $\mathcal{U}_i$ and record the resulting error syndromes.
Since the BB code is of CSS-type, $X$-type and $Z$-type errors can be decoded independently. In the following, we focus on the discussion on the case where a logical $Z$ basis state is prepared, and $X$-type errors are decoded.
We introduce the following notations:
\begin{itemize}
\item $S^{\mathcal{U}_{i}} \in\{0,1\}^{7t}$: the error syndrome of $\mathcal{U}_i$, obtained by measuring all $Z$-type check qubits over $t$ QEC cycles ;
\item $E_i=X(\alpha_i)$ (with $\alpha_i \in\{0,1\}^n$): the final Pauli error on the data qubits generated by $\mathcal{U}_i$ (we ignore Pauli $Z$ errors since they do not affect the logical $Z$ basis state) ;
\item $S_i^F = \alpha_i (H_Z)^{\top} \in\{0,1\}^7$: the error syndrome in the final cycle obtained from the data qubit readout ;
\item $S_i^L \in\{0,1\}^{4}$: the logical syndrome of the final error $E_i$. For a chosen basis of logical Pauli-$Z$ operators $\{ \bar{Z}_1, \bar{Z}_2, \bar{Z}_{3}, \bar{Z}_{4} \}$, the $j$-th entry of $S_i^L$ is $0$ if $E_i$ commutes with $\bar{Z}_j$ and $1$ if it anticommutes;
\item $\mathcal{D}$: the decoding matrix of size $M\times(7t+7)$, where the $i$-th row is given by $\left[S^{\mathcal{U}_{i}}, S_{i}^{F}\right]$ ;
\item $\mathcal{D}^{L}$ (of size $M\times4$): the $i$-th row of is $S_{i}^{L}$.
\end{itemize}
After performing the simulation to obtain $\mathcal{D}$ and $\mathcal{D}^{L}$, we merge the rows that correspond to identical syndrome triples $(S^{\mathcal{U}_{i}},S_{i}^{F},S_{i}^{L})$.
Let $M^{\prime}$ be the number of unique rows in $\mathcal{D}$ after this merging step.
We define $P=\{p_1,p_2,\ldots,p_{M^{\prime}} \}$, where $p_i$ is the error probability associated with the syndrome triple $(S_{i}^{\mathcal{U}},S_{i}^{F},S_{i}^{L})$. The value of $p_i$ is obtained by summing probabilities of all rows that were merged into the $i$-th row.

Consider a set of independent random variables $\xi=\{\xi_1, \xi_2, \ldots, \xi_{M^{\prime}}\} \in\{0,1\}^{M^{\prime}}$, where each $\xi_j$ takes the value $0$ with probability $1-p_j$ and $1$ with probability $p_j$.
Each configuration $(\xi_1, \xi_2, \ldots, \xi_M \in\{0,1\})$ corresponds to a unique faulty syndrome measurement circuit $\mathcal{U}_{\xi}$.
The final Pauli error on the data qubits generated by $\mathcal{U}_{\xi}$ is then given by: $E = \left(E_1\right)^{\xi_1}\left(E_1\right)^{\xi_2}\ldots \left(E_{M^\prime}\right)^{\xi_{M^\prime}}$. The error syndromes corresponding to $\mathcal{U}_{\xi}$ is obtained by summing (over $\mathbb{F}_2$) the rows of the decoding matrix $\mathcal{D}$ weighted by $\xi$:
\begin{equation}
{\left[\begin{array}{c}
S^{\mathcal{U}_\xi}, S^F_\xi
\end{array}\right] = \xi D = \sum_{i=1}^{M^\prime} \xi_i\left[\begin{array}{c}
S^{\mathcal{U}_i}, S_i^F
\end{array}\right] \quad(\operatorname{mod} \ 2)}
\end{equation}

With the error syndromes $\left(S^{\mathcal{U}},S^{F}\right)$ obtained from experiments or simulation, the BP-OSD decoder then seeks a solution of $\xi D^{\top}=\left(S^{\mathcal{U}},S^{F}\right)$ with the largest reliability using a greedy algorithm~\cite{Panteleev2021Degenerate}.

\section{Experimental details}
\begin{figure}[h]
\centering
\includegraphics[width=0.7\linewidth]{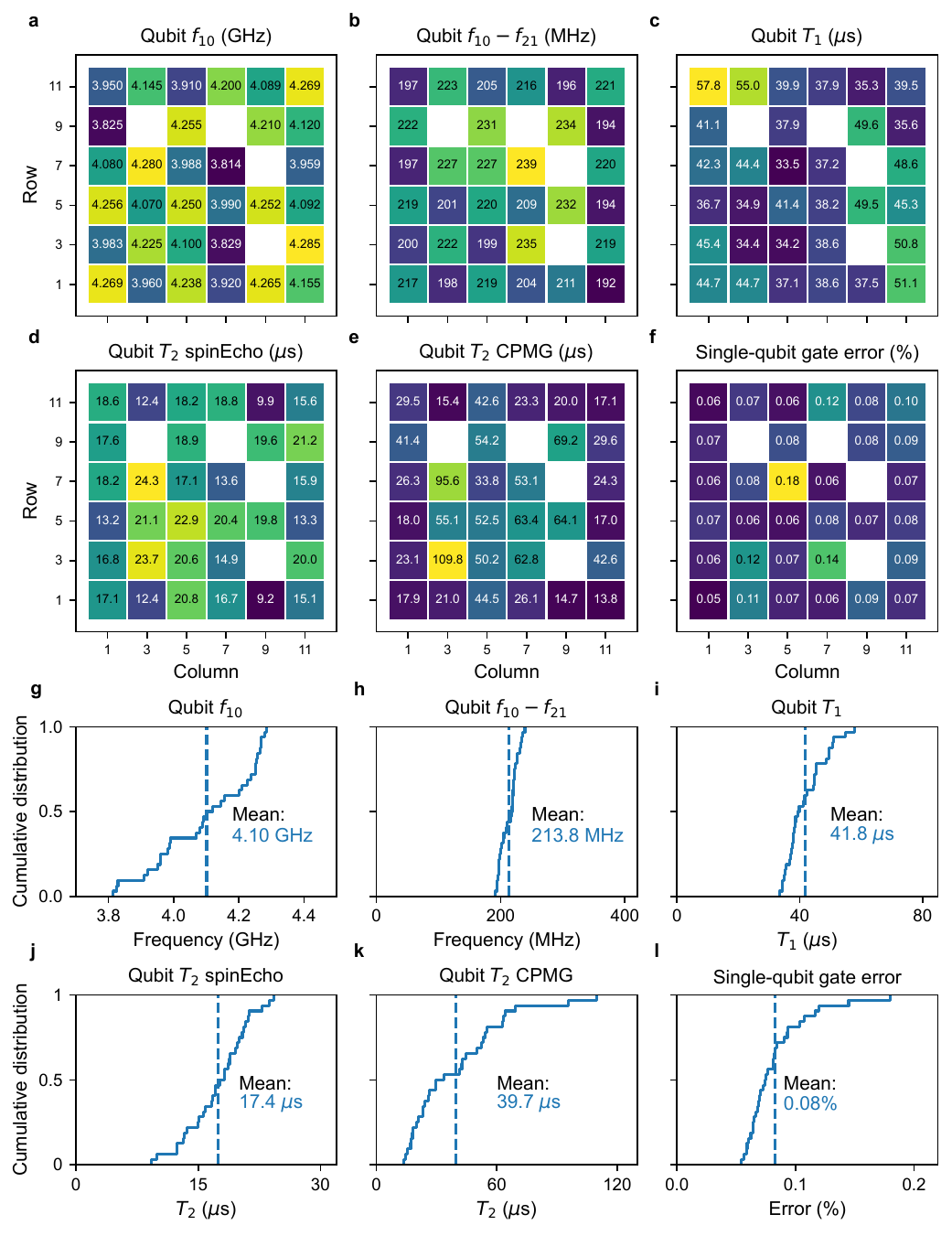}
\caption{
\textbf{
Heatmaps of single-qubit parameters and corresponding cumulative distributions.} 
\textbf{a,} Qubit idle frequency.
\textbf{b,} Qubit anharmonicity at the idle frequency.
\textbf{c,} Qubit relaxation time at the idle frequency.
\textbf{d,} Qubit pure dephasing time as measured using Hahn echo sequence.
\textbf{e,} Qubit pure dephasing time as measured by CPMG (Car-Purcell-Meiboom-Gill) sequence.
\textbf{f,} Pauli error of simultaneous single-qubit gates.
\textbf{g-l,} Corresponding cumulative distributions. 
Dashed lines indicate the mean values.
}
\label{fig:sq_info}
\end{figure}

\begin{figure}[h]
\centering
\includegraphics[width=0.8\linewidth]{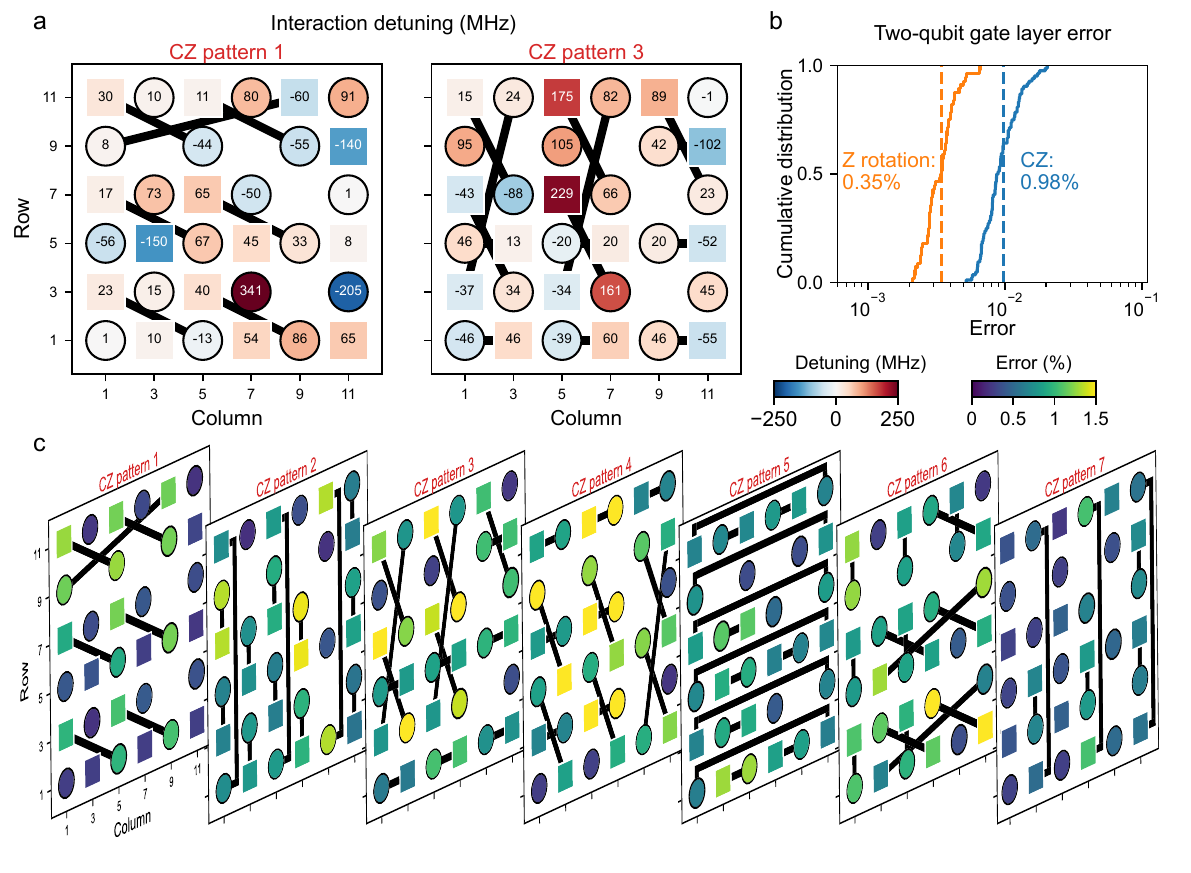}
\caption{
\textbf{Implementation of parallel CZ gates.}
\textbf{a,} Interaction detuning from the idle frequency of each qubit optimized during the execution of CZ pattern $1$ and $3$, corresponding to the first and third CZ layers in Fig.~\ref{fig:SM_circuit}. Note that the frequencies of qubits not involved in CZ gates are also optimized, resulting in a set of single-qubit $Z$ rotation gates applied simultaneously with the CZ gates.
In practice, we cancel out the effects of these additional $Z$ rotations with virtual $Z$ gates.
\textbf{b,} Cumulative distributions of $Z$ rotation errors and simultaneous CZ gate errors, with dashed lines indicating the mean values. 
\textbf{c,} Measured errors of CZ gates and $Z$ rotations using simultaneous cross-entropy benchmarking, where each CZ pattern corresponds to a CZ layer in Fig.~\ref{fig:SM_circuit}.
}
\label{fig:algorithm_graph}
\end{figure}

\subsection{Device Information}
The Kunlun quantum processor used in this experiment consists of $32$ frequency-tunable transmon qubits and $84$ tunable couplers.
Each qubit is connected to a readout resonator for dispersive measurement, and a dedicated control line for microwave and flux tuning.
As illustrated in Fig.~1\textbf{a} of the main text, the qubits are arranged in a square lattice, among which $18$ are data qubits (labeled by $R/L$) and $14$ serve as check qubits (labeled by $Z/X$) for implementing the BB code.
Each check qubit is connected to six data qubits through tunable couplers with length scales from approximately $1$~mm to $6.5$~mm.
To realize the connectivity topology of the BB code (Fig.~1\textbf{a} of the main text) in a two-dimensional architecture, we use air-bridges for enabling crossover between couplers.

\subsubsection{Multi-length couplers}
A key experimental challenge for implementing qLDPC codes on superconducting processors is realizing long-range qubit connectivity, which goes beyond the conventional nearest-neighbor interaction architecture.
To realize qubit interactions across varying distances while minimizing unwanted crosstalk interactions, we adopt the tunable coupler design.
Each coupler is realized as a frequency-tunable transmon, mediating interactions between two distant qubits through capacitive coupling. 
By varying the coupler lengths, we establish a capacitance gradient, producing nonlinearities spanning from $-10$ MHz to $-170$ MHz.
To maintain consistent coupling tunability across all couplers despite this variation, we carefully engineer the qubit-coupler coupling capacitance and the Josephson junction resistances of the couplers to compensate for the capacitance differences.

In Kunlun, we use air bridges to overcome the constraints of two-dimensional architectures in constructing high-connectivity quantum processors.
First, we incorporate multiple air bridges into the capacitance of the couplers, effectively forming a quasi-three-dimensional structure in regions where couplers intersect.
In addition, we use air bridges to connect the grounds that are separated by couplers and control lines for maintaining a uniform ground reference.

\subsubsection{Measurement setup}
The device contains six readout lines, each coupled to five or six readout resonators. 
To minimize crosstalk during readout, the resonators on each line are frequency-multiplexed with a frequency separation of approximately $100$~MHz, enabling simultaneous measurements with minimal interference. 
The qubit-resonator coupling strengths are designed to be approximately $200$~MHz for check qubits and $80$~MHz for data qubits.
In addition, we implement a Purcell filter on each readout line to suppress qubit decay caused by the Purcell effect while maintaining a high signal-to-noise ratio.

\subsection{Quantum operations}
A full cycle of syndrome measurement consists of seven layers of two-qubit CZ gates, interspersed with Hadamard and echo gates, followed by the measurement of check qubits.
The performance of these quantum operations is critical for implementing quantum error correction codes.
In this subsection, we summarize the parameters and performances of these quantum operations.

\subsubsection{Single- and two-qubit gates}
\label{sec:CZ_gate}
Arbitrary single-qubit gates are implemented by combining $XY$ rotations with virtual-$Z$ gates~\cite{Xu2024NonAbelian}, where $XY$ rotations are implemented using $30$-ns-long microwave pulses resonant with the $|0\rangle\leftrightarrow|1\rangle$ transition frequencies.
To realize high-fidelity parallel single-qubit gates, we carefully optimize the idle frequencies at which the $XY$ pulses are applied, taking into account decoherence times and parasitic couplings between qubits. In addition, we use an active microwave compensation technique to eliminate microwave crosstalk~\cite{Ren2022Experimental}.
Fig.~\ref{fig:sq_info}\textbf{a}-\textbf{f} summarize the qubit idle frequencies, nonlinearities, and coherence times (including energy relaxation and dephasing times) of all qubits.
Fig.~\ref{fig:sq_info}\textbf{g}-\textbf{l} present the Pauli errors of the parallel single-qubit gates characterized using simultaneous cross-entropy benchmarking (XEB).

Two-qubit CZ gates are realized by bringing $|11\rangle$ and $|02\rangle$ (or $|20\rangle$) of qubit pairs in near resonance at the respective interaction frequencies and activating the coupling for a duration of $95$~ns, with two additional $5$-ns buffers applied at the beginning and end. 
As shown in Fig.~\ref{fig:SM_circuit}, a cycle of the syndrome measurement circuit for implementing the BB code contains seven layers of CZ gates. 
For each CZ layer composed of $7$ or $14$ parallel gates, we optimize the interaction frequencies to maximize overall gate fidelity, taking into account qubit coherence times, pulse shaping, and parasitic couplings between non-interacting qubits~\cite{Xu2024NonAbelian}.
The optimized interaction frequencies for the first and third CZ layers are presented in Fig.~\ref{fig:algorithm_graph}\textbf{a}. 
The Pauli error rates of the parallel CZ gates, characterized using simultaneous XEB, are shown in Fig.~\ref{fig:algorithm_graph}\textbf{b},\textbf{c}.

\subsubsection{Readout}
To maximize both the accuracy and speed of readout in our measurement setup, we apply a $520$-ns readout pulse for check qubits ($890$-ns for data qubits), followed by a $400$-ns buffer for photon decay. 
To achieve optimal readout performance, we apply a mode-matched filter to the response signal. This is based on time-resolved measurements of the resonator response to the readout pulse during the demodulation of readout pulses~\cite{Heinsoo2018Rapid}.

\begin{figure}[t]
\centering
\includegraphics[width=0.7\linewidth]{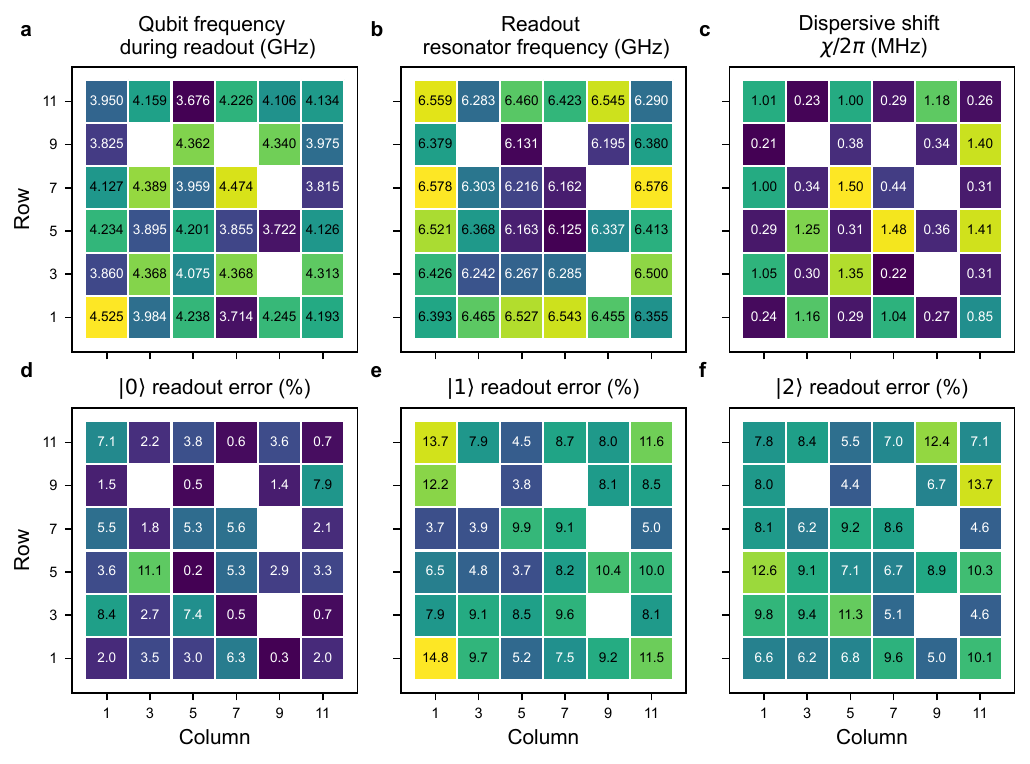}
\caption{
\textbf{Heatmaps of readout information.} 
\textbf{a,} Qubit frequencies during the readout procedure.
\textbf{b,} Frequencies of the readout resonators.
\textbf{c,} Measured dispersive shift $\chi$ for each qubit.
\textbf{d-f,} Readout errors of the qubit $\ket{0}$, $\ket{1}$, and $\ket{2}$ states.
}
\label{fig:readout}
\end{figure}

During the dispersive readout, qubit frequencies change due to the AC-Stark effect and may collide with others, causing unwanted readout crosstalk. To eliminate such crosstalk and realize high-fidelity simultaneous mid-circuit readout, we optimize the qubit frequencies by applying flux pulses during the implementation of readout pulses.
The optimized qubit frequencies during readout, the frequencies of readout resonators and the dispersive shift $\chi$, are shown in Fig.~\ref{fig:readout}.
To characterize the simultaneous readout fidelity, we randomly prepare each qubit in the $|0\rangle$, $|1\rangle$, or $|2\rangle$ states and measure the assignment errors. The averaged readout errors for three-state measurements are shown in Fig.~\ref{fig:readout}.

\subsection{Characterization of stabilizer measurements}

\begin{figure}[t]
\centering
\includegraphics[width=1\linewidth]{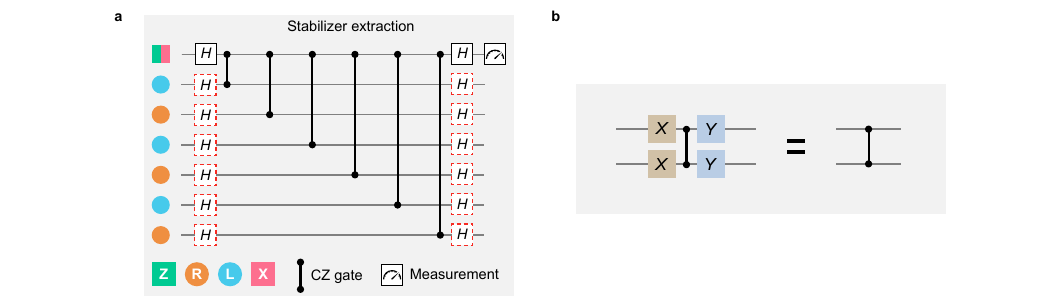}
\caption{
\textbf{Single stabilizer measurement.}
\textbf{a,}
Quantum circuit for extracting a weight-$6$ $X$-type or $Z$-type stabilizer, without additional Pauli gates inserted between CZ layers.
The circuit applies sequential CZ gates between the check qubit and each of the six data qubits. The Hadamard gates (marked by red dashed squares) are applied on the data qubits only when measuring an $X$-type stabilizer.
\textbf{b,}
Equivalent quantum circuits illustrating that inserting Pauli $X$ gates and $Y$ gates to the left and right of a CZ gate, respectively, leaves the CZ gate unchanged.
}
\label{fig:stabilizer_no_protect}
\end{figure}

With well-optimized quantum gates and readout, we characterize the measurements of $14$ individual weight-$6$ stabilizers for the implemented BB code.
Fig.~\ref{fig:stabilizer_no_protect}\textbf{a} shows the original quantum circuit for single stabilizer measurement. To mitigate dephasing during two-qubit gates, we introduce Pauli gates between adjacent CZ gates, exploiting the identity shown in Fig.~\ref{fig:stabilizer_no_protect}\textbf{b}.
The final circuit is illustrated in Fig.~2\textbf{c} of the main text.

\begin{figure}[h]
\centering
\includegraphics[width=1\linewidth]{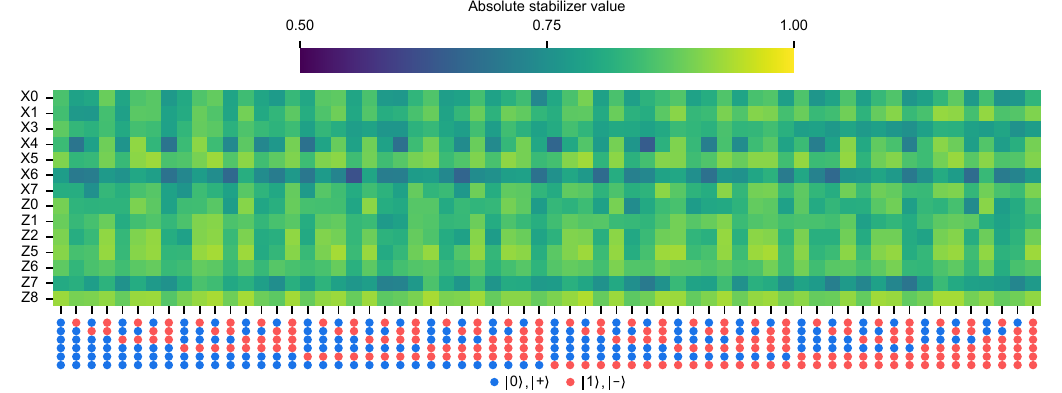}
\caption{
\textbf{Characterization of stabilizer measurements for $\boldsymbol{14}$ weight-$\boldsymbol{6}$ stabilizers.} 
The $x$-axis labels denote the input basis states of the six data qubits involved in each stabilizer, represented by color-coded circles: blue for $\ket{0}$ (or $\ket{+}$) and red for $\ket{1}$ (or $\ket{-}$).
}
\label{fig:stabilizer}
\end{figure}

For each stabilizer, we sequentially prepare its six coupled data qubits into all $2^6=64$ basis states and then execute the stabilizer extraction circuit for $3,000$ repetitions per basis.
A basis state is a product state composed of $|0\rangle$ and $|1\rangle$ for a $Z$-type stabilizer and $|+\rangle$ and $|-\rangle$ for an $X$-type stabilizer.
We compute the average of the measured stabilizer value for each basis and each stabilizer, with the results shown in Fig.~\ref{fig:stabilizer}.
On average, the experimental error for weight-$6$ stabilizer measurements is $8.1\pm2.8\%$.

\subsection{Dephasing error mitigation}
\begin{figure}[t]
\centering
\includegraphics[width=1\linewidth]{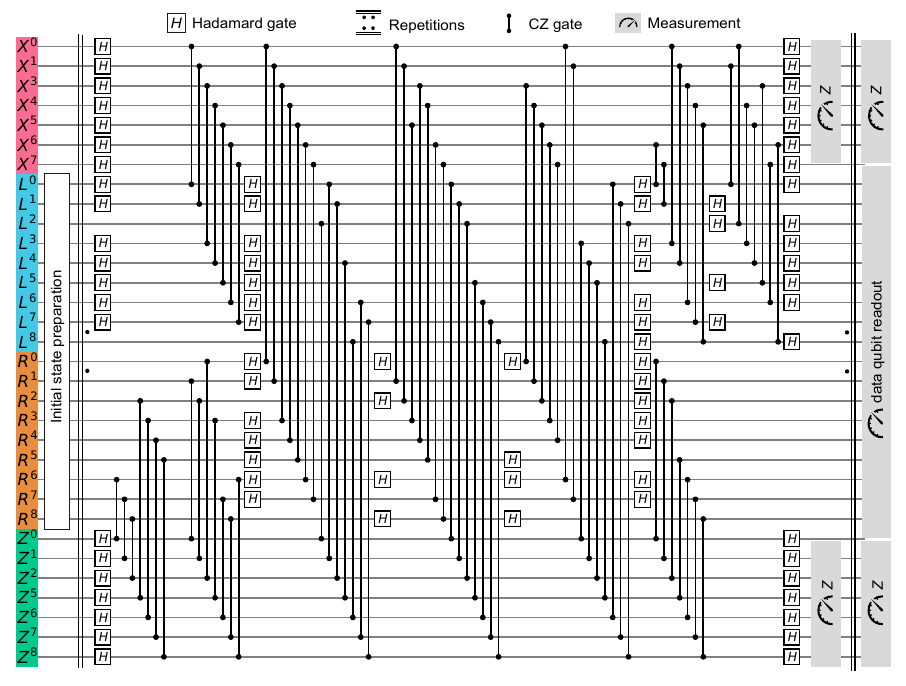}
\caption{
\textbf{Quantum circuit used for periodic stabilizer measurements of the $\boldsymbol{[[18,4,4]]}$ BB code, without additional Pauli gates inserted between CZ layers and without dynamical decoupling applied.}
}
\label{fig:circuit_without_protect}
\end{figure}

\begin{figure}[h]
\centering
\includegraphics[width=1\linewidth]{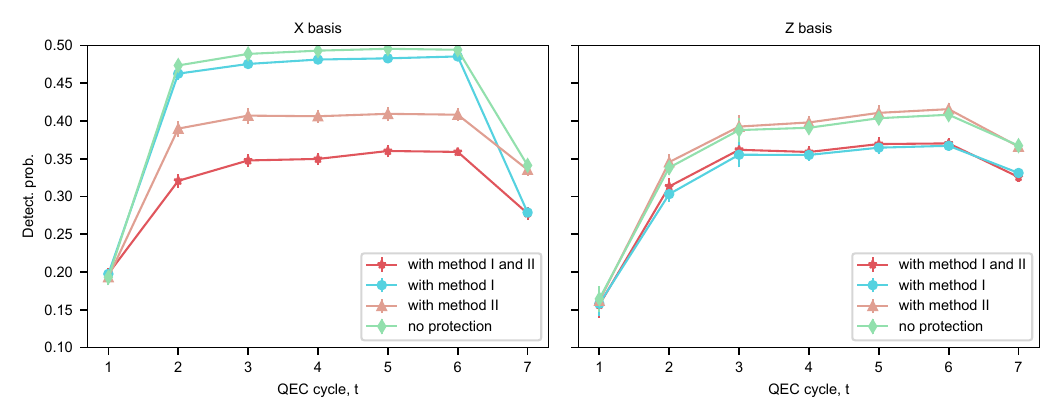}
\caption{
\textbf{Detection probabilities for both $X$-type and $Z$-type stabilizers under different dephasing mitigation methods.}
Method I refers to the insertion of Pauli gates around CZ layers, while method II corresponds to the application of dynamical decoupling gates during check qubit readout.
}
\label{fig:mean_detection_errors}
\end{figure}

In our experiment, data qubit decoherence is a significant source of errors, with dephasing being the primary mechanism.
Without the application of echos or dynamical decoupling (DD) techniques, the syndrome measurement circuit for the implemented BB code is shown in Fig.~\ref{fig:circuit_without_protect}.
To mitigate the dephasing effects, we adopt two complementary methods in our experiment.
First, we strategically insert Pauli $X$ or $Y$ gates around each CZ gate to serve as echo pulses.
To preserve the overall circuit unitary, these echo pulses are supplemented with compensatory single-qubit rotations in the final layer.
Second, during the $920$-ns-long readout of check qubits, we uniformly apply ten Pauli $X$ gates to each data qubit, serving as DD pulses to suppress low-frequency dephasing noise.
To evaluate the effectiveness of the two dephasing mitigation methods, we compare the detection probabilities for the BB code over seven syndrome cycles, with and without applying the two methods.
As shown in Fig.~\ref{fig:mean_detection_errors}, we observe a clear reduction in detection probabilities when either method is applied, with the best performance achieved when both methods are combined.
In addition, we find that dephasing leads to a more pronounced performance degradation in $X$-basis experiments compared to $Z$-basis experiments.

\subsection{Leakage rejection}
\begin{figure}[t]
\centering
\includegraphics[width=0.85\linewidth]{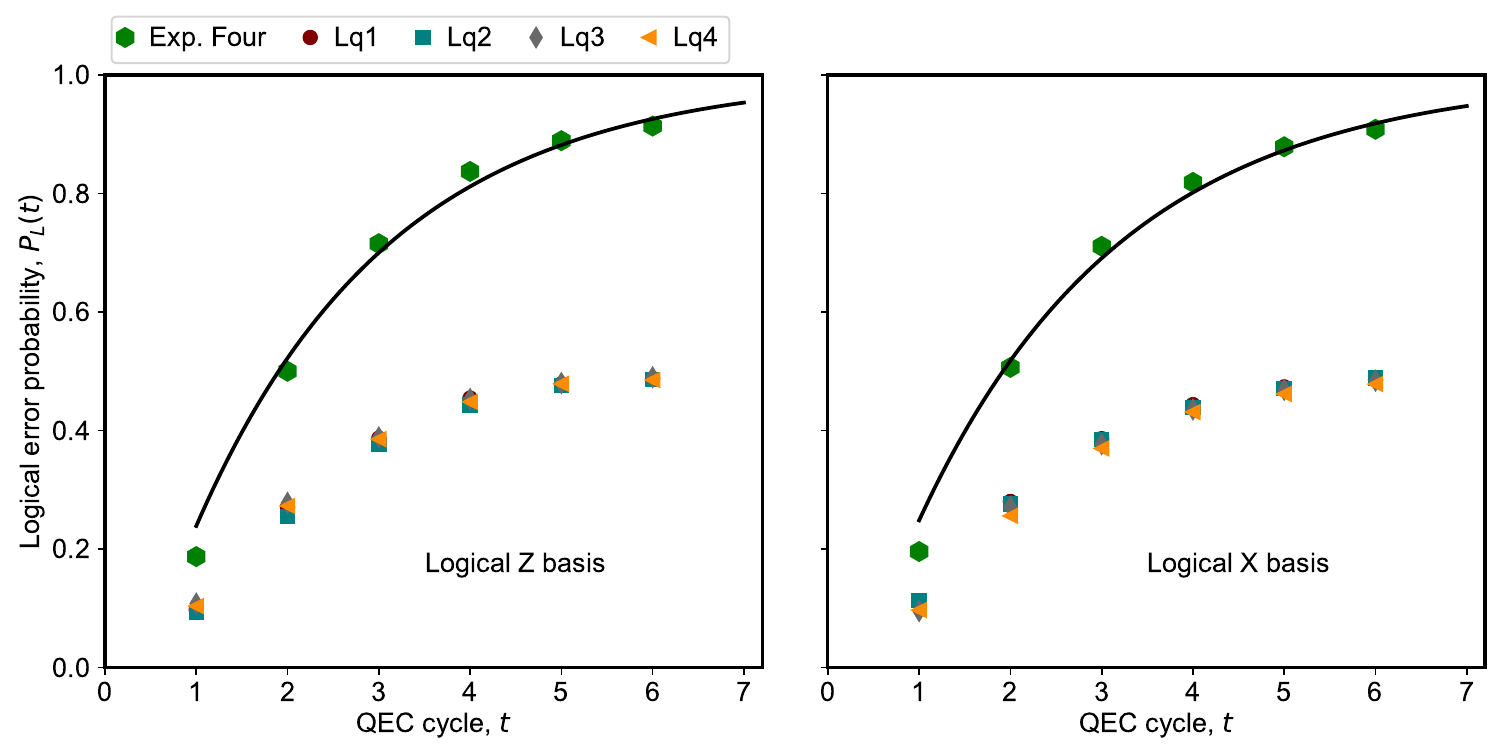}
\caption{
\textbf{Logical performance for the BB code $\boldsymbol{[[18,4,4]]}$ without leakage rejection in experiments.}
Each individual data point represents $40,000$ experimental instances. Solid line: fit to the experimental data from $t = 1$ to $6$.
}
\label{fig:leakage_Logical_error_18_4_4}
\end{figure}

Leakage into the non-computational subspace introduces long-lived errors that propagate across consecutive syndrome cycles, thereby increasing the complexity of decoding and degrading logical performance.
To detect leakage, we employ a three-state readout scheme capable of distinguishing the ground state $\ket{0}$, the first excited state $\ket{1}$, and the second excited state $\ket{2}$.
Any experimental instance is rejected if leakage to the $\ket{2}$ state is detected on any qubit.
In Fig.~\ref{fig:leakage_Logical_error_18_4_4}, we present the logical error probabilities of the implemented BB code without excluding experimental instances in which leakage is detected.
We fit to obtain the logical error rate per qubit per cycle of $10.99\%$ for $Z$ bases and $10.51\%$ for $X$ bases.
Compared to the leakage-rejected results shown in Fig.~3\textbf{a},\textbf{b} of the main text, the absence of leakage rejection leads to a clear increase in logical error per qubit per cycle, with an increase of $1.84\%$ for both $Z$ bases and $X$ bases.

The fraction of retained experimental instances after excluding leakage, as a function of the number of syndrome cycles, is listed in Tab.~\ref{tab:leakage_rejection}.
In addition, we estimate the corrected probability of leakage into the $\ket{2}$ state by applying readout correction.
The leakage rate per cycle is then determined by averaging the increase in the corrected leakage population over six syndrome cycles, as shown in Tab.~\ref{tab:leakage_rate}.
We find that the probability of having detected a leakage event on any of the 18 data qubits is approximately $0.127(3)$, while for the 14 check qubits, it is about $0.082(6)$.
The average leakage rate per cycle for one data qubit is $0.007(5)$, slightly higher than the $0.006(1)$ for one check qubit. 
This difference is attributed to the frequency arrangement of the qubits, where data qubits are operated at higher frequencies during CZ gates, leading to a higher leakage rate.

\begin{table}[h]
    \centering
    \renewcommand{\arraystretch}{1.2}
    \begin{tabular}{c c}
        \toprule
        \textbf{Cycle index} & \textbf{Retained data fraction} \\
        \midrule
        1  & 31.5~\% \\
        2  & 15.7~\% \\
        3  & 7.4~\% \\
        4  & 3.9~\% \\
        5  & 2.2~\% \\
        6  & 1.0~\% \\
        \bottomrule
    \end{tabular}
    \caption{\textbf{Retained data fraction as a function of the number of syndrome cycles after leakage rejection.}}
    \label{tab:leakage_rejection}
\end{table}

\begin{table}[t]
    \centering
    \renewcommand{\arraystretch}{1.2}
    \begin{tabular}{l c}
        \toprule
        \textbf{Type} & \textbf{Leakage rate per cycle} \\
        \midrule
        Data qubits only   & 0.127(3) \\
        Check qubits only & 0.082(6) \\
        One data qubit & 0.007(5) \\
        One check qubit & 0.006(1) \\
        \bottomrule
    \end{tabular}
    \caption{\textbf{Leakage rates per cycle after readout correction.}}
    \label{tab:leakage_rate}
\end{table}

\section{Data analysis}
\subsection{Converting measurement outcomes into error detection results}
\begin{figure}[h]
\centering
\includegraphics[width=0.85\linewidth]{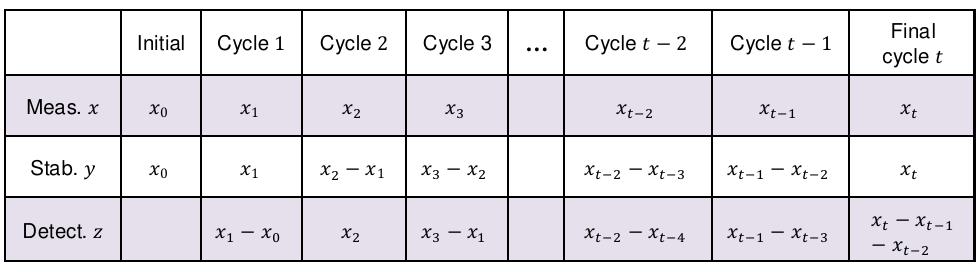}
\caption{
\textbf{
Conversion of measurement outcomes of check qubits into error detection results.} 
We denote the check qubit measurement outcome, the stabilizer value, and the error detection result as $x$, $y$, and $z$, respectively.
The actual stabilizer value is given by $y_j=(x_j-x_{j-1}) \mod 2$ except for the first and final cycles. The error detection result is then computed as $z_j=(y_j-y_{j-1}) \mod 2$.
}
\label{fig:meas_stab_detect}
\end{figure}

In our experiment, we convert measurement outcomes of check qubits into error detection results. As illustrated in Fig.~\ref{fig:meas_stab_detect}, this process involves two steps.
First, due to the absence of check qubit resets, we must recover the actual value of each stabilizer from the corresponding check qubit measurement outcome. Specifically, except for the first and final cycles, suppose the readout outcome of a check qubit in the $j$-th cycle is $x_j$, and the readout outcome in the preceding $j-1$-th cycle is $x_{j-1}$. If $x_{j-1}=0$, the stabilizer value in the $j$-th cycle is simply $y_{j}=x_{j}$. However, if $x_{j-1}=1$, the stabilizer value is given by $y_{j} = (x_{j}-1) \mod 2$. Thus, the actual stabilizer value should be computed as $y_{j}=(x_{j+1}-x_{j}) \mod 2$.
For the first cycle and final cycles, the stabilizer values are directly given by $y_{j}=x_{j}$, since there is no preceding cycle for the first cycle, and the final cycle parity outcomes are obtained directly from data qubit measurements.
Second, we convert the stabilizer values into error detection events. An error detection event is recorded if a stabilizer value differs from the corresponding one in the preceding cycle. Consequently, the error detection result in the $j$-th cycle, denoted as $z_j$, is given by $z_j=(y_j-y_{j-1}) \mod 2$.

\end{document}